\documentclass[11pt]{article}

\usepackage{caption}
\usepackage{etoolbox}
\usepackage{mathrsfs}
\usepackage{amsmath}
\usepackage{amssymb}
\usepackage[colorlinks=true,urlcolor=blue,linkcolor=black,citecolor=black]{hyperref}
\hypersetup{pdfstartview={XYZ null null 1},
            pdftitle={Extended Particles and the Exterior Calculus},
            pdfauthor={R W Tucker},
            }
\usepackage{tocloft}
\usepackage{graphicx}

\usepackage{array}
\newcolumntype{L}[1]{>{\raggedright\let\newline\\\arraybackslash\hspace{0pt}}m{#1}}
\newcolumntype{C}[1]{>{\centering\let\newline\\\arraybackslash\hspace{0pt}}m{#1}}

\renewcommand{\arraystretch}{1.7}

\def\={\;\;=\;\;}                               % Equals sign with extra spacing
\def\df{\frac{}{}}                              % Dummy fraction
\renewcommand{\mathbf}[1]{\boldsymbol{#1}}
\def\diff#1#2{\frac{d#1}{d#2}}                  % Differentiation
\def\pdiff#1#2{\frac{\partial#1}{\partial#2}}   % Partial differentiation
\def\({\left(}                                  % Large left bracket (
\def\){\right)}                                 % Large right bracket )
\def\[{\left[}                                  % Large left bracket [
\def\]{\right]}                                 % Large right bracket ]
\def\real{\mathbb{R}}
\def\wh#1{\widehat{#1}}
\def\wt#1{\widetilde{#1}}
\def\tensor{\otimes}
\def\w{\wedge}
\def\man#1{\mathcal{#1}}
\def\choose#1#2{\(	\begin{smallmatrix}	
						#1 \\
						#2
					\end{smallmatrix}\) }
\def\g{g}

\def\qquadand{\qquad\textrm{and}\qquad}

\def\bibtitle#1{\mbox{\normalsize \textit{#1}}}

\newcounter{defncounter}
\numberwithin{defncounter}{section}

\numberwithin{equation}{section}
\def\real{\mathbb{R}}
\renewcommand{\thesection}{\Roman{section}} 
\def\bvec{e}
\def\dbasis{\varepsilon}
\def\SS{\mathcal{P}}
\def\M{\mathcal{M}}
\def\N{\mathcal{N}}
\def\U{\mathcal{U}}

\begin{document}

\title{	\fbox{\parbox[3cm]{0.6\textwidth}{
				\centering \quad \\[0.4cm]
				{\huge Extended Particles and the Exterior Calculus} \\[1cm] 
				{\Large Rutherford Laboratory \\
				Chilton, Didcot, Oxon, OX11 0QX} \\[0.5cm] }} \\[2cm]
		}
\author{R. W. Tucker \\[0.2cm]
		University of Lancaster and Daresbury Laboratory\\[0.4cm]}
		
\date{February 1976 \\[0.4cm]
	RL-76-022}

\maketitle
\thispagestyle{empty}
\newpage

\thispagestyle{empty}
\quad \\[4cm]
\begin{center}
	Notes based on a series of invited \\
	lectures given at the Rutherford Laboratory \\
	February 1976
\end{center}
\newpage

\tableofcontents
\newpage

\addtocontents{toc}{\protect\thispagestyle{empty}}

\setcounter{page}{1}

%%%%%%%%%%%%%%%%%%%%%%%%%%%%%%%%%%%%%%%%%%%%%%%%%%%%%%%%%%%%%%%%%%%%%%%%%%%%%%
\section*{Introduction}
\addcontentsline{toc}{section}{\protect\numberline{}Introduction}
%%%%%%%%%%%%%%%%%%%%%%%%%%%%%%%%%%%%%%%%%%%%%%%%%%%%%%%%%%%%%%%%%%%%%%%%%%%%%%
The purpose of these lectures is primarily two-fold:
\begin{enumerate}
	\item	To discuss the classical theory of free point particles, free strings and free membranes from a unified viewpoint.
	\item	To present in the process of doing this the rudiments of an intrinsic geometrical calculus that the author has found of immense value in investigating these systems. 
\end{enumerate}
The geometry of submanifolds and the calculus of exterior differential forms are widely exploited in many branches of mathematics and gravitational physics \cite{Gravity}. The interest in classically extended relativistic systems as potential models for elementary particles raises the possibility of their usefulness as tools in high energy physics. \\

The essential characteristics of a geometric calculus is its independence of co-ordinate representations. In the language of extended particles this means that the theory can be discussed in a reparameterisation or gauge invariant way. One can consequently concentrate on the intrinsic aspects of the system. Although we shall not discuss interactions in any detail in these notes, the formalism certainly suggests how external fields (or potentials) may be coupled in a way that does not spoil the reparameterisation invariance. \\

It will be shown that the equations of motion arise in a very simple manner from a principle of stationary action and furthermore the boundary conditions for finite systems are derived in a gauge invariant way. Momenta are naturally introduced and the primary constraints that exist in a Hamiltonian description follow simply. The calculations can proceed in an index-free manner until components are required. It is at this stage that one can if one desires impose gauge conditions and remove non-independent degrees of freedom. \\

The methods can be applied in a space-time of any dimension and metric. \\

There exist at present a large number of excellent reviews on the classical and quantum theories of relativistic strings \cite{Strings}. These notes will not be concerned with the intricacies and problems of consistent quantisation or with the rules that exist for the construction of scattering amplitudes. However, it is hoped that these notes will help elucidate the basic essentials of string dynamics and provide a means of naturally generalising to systems with other degrees of freedom. \\

Section~\ref{Sect:ExtCal} establishes the basic formalism used in the rest of the notes. Some concepts are mentioned only briefly but the interested reader will find ample information in the literature quoted \cite{ExtCalculus}. The reader is warned that mathematical notation is by no means uniform and in some cases its economy seems designed to promote elegance rather than to assist in computation. This first section is hopefully self contained but in the process of distillation all proofs have been relegated to the literature. Section~\ref{Sect:RelPoint} formulates the familiar relativistic point particle in the language of differential forms. In this case electromagnetic and gravitational couplings are explicitly accommodated. Section~\ref{Sect:RelString} generalises the procedure to relativistic strings and the question of fixing a suitable gauge is discussed in some detail. The last section discusses some recent work concerned with a generalisation to relativistic membranes. 
\newpage

\addtocontents{toc}{\protect\vspace{10pt}}

%%%%%%%%%%%%%%%%%%%%%%%%%%%%%%%%%%%%%%%%%%%%%%%%%%%%%%%%%%%%%%%%%%%%%%%%%%%%%%
\section{The Exterior Calculus}\label{Sect:ExtCal}
%%%%%%%%%%%%%%%%%%%%%%%%%%%%%%%%%%%%%%%%%%%%%%%%%%%%%%%%%%%%%%%%%%%%%%%%%%%%%%
The formalism to be described is approached only in the language of finite-dimensional vector spaces. We recall that these are composed of abstract elements which, with the operations of addition and multiplication by scalars, may be thought of as points or vectors. The requisite associativity and commutativity of addition, the existence of a zero vector and the distributive property of scalar multiplication are all satisfied by elements composed of ordered set of $n$ real numbers. These elements may be taken to compose the fundamental space $\real^{n}$ (sometimes called arithmetic $n$-space). A standard basis for this vector space will be denoted by $\wh{\bvec}_{1}, \,\wh{\bvec}_{2}, \ldots, \wh{\bvec}_{n}$ where $\wh{\bvec}_{j}$ is the ordered $n$-tuple $(0,0,\ldots,1,0,\ldots, 0)$ with unity in the $j$-th position. A general vector when regarded as a geometric entity will be denoted by letters such as $U$ or $V$ without any suffixes. In terms of a particular basis with vectors $\{\bvec_{i}\}$, we introduce real components $V^{i}$ of $V$ by
\begin{eqnarray}\label{vect_comp}
	V	&=& V^{i}\bvec_{i}	\qquad i\;=\; 1,2,\ldots, n
\end{eqnarray}
and the usual summation convention is adopted. A vector space becomes Euclidean as soon as we introduce a real bilinear symmetric scalar product $U\cdot V$ of two vectors such that $V\cdot V>0$ if $V$ is non-zero and with orthonormal basis vectors satisfying $\bvec_{i}\cdot\bvec_{j} = \delta_{ij}$, a norm is defined by
\begin{eqnarray*}
	|V| &=& \(V\cdot V\)^{\frac{1}{2}} \;=\; \( \df (V^{1})^{2} + (V^{2})^{2} + \cdots + (V^{n})^{2} \)^{\frac{1}{2}}.
\end{eqnarray*}
We can form vector spaces of tensors on a given vector space by means of the tensor product $\tensor$. However, there exist a class of tensors with antisymmetric components that play an important role in what follows. Given two tensors $S$ and $T$, we can define their \textit{Grassman product}
\begin{eqnarray*}
	S \w T &=& \man{A}(S \tensor T)
\end{eqnarray*}
where $\man{A}$ applied to any tensor antisymmetrises its components (and supplies a conventional normalization factor). This product can be defined independently of the tensor product if we introduce the Grassman product $\w$ to be anti-commutative (but associative and distributive)
\begin{eqnarray}\label{def_G_anti}
	\bvec_{i} \w \bvec_{j} &=& -\bvec_{j} \w \bvec_{i}
\end{eqnarray}
for any pair of basis vectors. \\

On an $n$-dimensional vector space, the $p-$multivectors are constructed from linear combinations of $\bvec_{i_{1}} \w \bvec_{i_{2}} \w \cdots \w \bvec_{i_{p}}$ with real coefficients. Clearly, only $n!/[p!(n-p)!]$ of these are independent and they may be chosen as a basis in the space $\Lambda_{p}$ with their indices ordered so that
\begin{eqnarray*}
	i_{1} \;<\; i_{2} \;<\; \cdots \;<\; i_{p}.
\end{eqnarray*}
Some examples are shown in table~\ref{table:Lam_p}. \\

\begin{table}[!h]
	\centering\small
	\begin{tabular}{|C{2cm}||C{2.4cm}||C{2cm}||C{2cm}||C{2.5cm}|}
		\hline 
		\bf Dimension of vector space & \bf Vector basis for $\boldsymbol{\Lambda_{1}}$ & \bf 2-vector basis for $\boldsymbol{\Lambda_{2}}$ & \bf 3-vector basis for $\boldsymbol{\Lambda_{3}}$ & \bf 4-vector basis for $\boldsymbol{\Lambda_{4}}$ \\
		\hline \hline
		2 	& $\bvec_{1}, \bvec_{2}$						& $\bvec_{1} \w \bvec_{2}$	& 							& \\[0.2cm]
		\hline\hline
		3 	& $\bvec_{1}, \bvec_{2}, \bvec_{3}$				& $\bvec_{1} \w \bvec_{2}$  & $\bvec_{1} \w \bvec_{2} \w \bvec_{3}$	& \\
			&												& $\bvec_{1} \w \bvec_{3}$	& 							& \\	
			&												& $\bvec_{2} \w \bvec_{3}$	& 							& \\[0.2cm]
		\hline\hline
		4 	& $\bvec_{1}, \bvec_{2}, \bvec_{3}, \bvec_{4}$	& $\bvec_{1} \w \bvec_{2}$	& $\bvec_{1} \w \bvec_{2} \w \bvec_{3}$		& $\bvec_{1} \w \bvec_{2} \w \bvec_{3} \w \bvec_{4}$ \\
							&								& $\bvec_{1} \w \bvec_{3}$	& $\bvec_{1} \w \bvec_{2} \w \bvec_{4}$	& \\
							&								& $\bvec_{1} \w \bvec_{4}$	& $\bvec_{1} \w \bvec_{3} \w \bvec_{4}$	& \\
							&								& $\bvec_{2} \w \bvec_{3}$	& $\bvec_{2} \w \bvec_{3} \w \bvec_{4}$	& \\
							&								& $\bvec_{2} \w \bvec_{4}$	& 							& \\
							&								& $\bvec_{3} \w \bvec_{4}$	& 							& \\[0.2cm]
		\hline 
	\end{tabular}
	\caption{Table showing basis vectors for $\Lambda_{p}$ with underlying vector spaces of various dimension.}
	\label{table:Lam_p}
\end{table}

In terms of components, if
\begin{eqnarray*}
	U \;=\; U^{i}\bvec_{i} \qquadand V \;=\; V^{i}\bvec_{i},
\end{eqnarray*}
then
\begin{eqnarray*}
	U \w V \;=\; \sum_{i,j}^{n} U^{i}V^{j}\bvec_{i} \w \bvec_{j} \;=\; \sum_{i<j}^{n} \( U^{i}V^{j} - U^{j}V^{i} \)\,\bvec_{i} \w \bvec_{j}.
\end{eqnarray*}
A general $p$-vector $W$ may be written
\begin{eqnarray}
	W &=& \sum_{i_{1} < i_{2} < \cdots \;<\; i_{p}.}^{n} W^{i_{1}i_{2}\cdots i_{p}} \, \bvec_{i_{1}} \w \bvec_{i_{2}} \w \cdots \w \bvec_{i_{p}}
\end{eqnarray}
and the $W^{i_{1}i_{2}\cdots i_{p}}$ are its components. If a general $p-$vector $W$ is multiplied by a $q-$vector $Y$ then from (\ref{def_G_anti})
\begin{eqnarray*}
	W \w Y &=& (-1)^{pq}\, Y \w W.
\end{eqnarray*}
Thus, although $\bvec_{j} \w \bvec_{j}$ is always zero, $W \w W$ need not be. The set of linear functions that map vectors into numbers can be added and multiplied by scalars and so form a vector space which is said to be the dual space of the former. The real valued linear functions are called covectors. Except in those cases in which we wish to identify with a common nomenclature, they will be denoted by Greek letter. Thus $\alpha$ is the map
\begin{eqnarray*}
	\begin{split}
		\alpha: \real^{n}	&\longrightarrow\;	\real \\
				V			&\longmapsto\;		\alpha(V).
	\end{split}
\end{eqnarray*} 
In some cases the notation $\alpha(V)$ is replaced by $\alpha V$, especially where a profusion of brackets would lead to confusion. If $\bvec_{1},\ldots, \bvec_{n}$ is a vector basis in $n$-dimensional $\Lambda_{1}$, the dual basis $\dbasis^{1},\ldots\,\dbasis^{n}$ is the set of elements defined by
\begin{eqnarray}\label{def_covec_basis}
	\dbasis^{i}(\bvec_{j}) &=& \delta^{i}_{j} \qquad i,j\;=\; 1,2,\ldots, n.
\end{eqnarray}
Once a basis is chosen the two spaces are isomorphic. Any covector $\alpha$ can be expanded as
\begin{eqnarray*}
	\alpha &=& \alpha_{i}\dbasis^{i}
\end{eqnarray*}
in terms of its real components $\alpha_{i}$. The evaluation of this covector on the vector $V$ of (\ref{vect_comp}) is the number
\begin{eqnarray*}
	\alpha(V) &=& \alpha_{i}V^{i}
\end{eqnarray*}
using the standard summation convention over repeated suffices. Multi-covectors are defined as linear mappings of multi-vectors into real numbers. If $\Lambda^{p}$ and $\Lambda_{p}$ denote the $\choose{n}{p}-$dimensional vector spaces of $p-$covectors and $p-$vectors respectively, a general multi-covector $\beta$ is the mapping
\begin{eqnarray*}
	\begin{split}
		\alpha: \Lambda_{p}	&\longrightarrow\;	\real \\
				V			&\longmapsto\;		\beta(V).
	\end{split}
\end{eqnarray*} 
In terms of the dual basis of $\Lambda^{p}$ composed of vectors
\begin{eqnarray*}
	\dbasis^{i_{1}} \w \dbasis^{i_{2}} \w \cdots \dbasis^{i_{p}} \qquad i_{1} \;<\; i_{2} \;<\; \cdots \;<\; i_{p},
\end{eqnarray*}
such a general $p-$covector has the expansion
\begin{eqnarray*}
	\beta &=& \sum_{i_{1} < i_{2} < \cdots \;<\; i_{p}}^{n}  \beta_{i_{1}i_{2}\cdots i_{p}}\, \dbasis^{i_{1}} \w \dbasis^{i_{2}} \w \cdots \w \dbasis^{i_{p}}
\end{eqnarray*}
and the numbers $\beta_{i_{1}i_{2}\cdots i_{p}}$ constitute its components in this basis. In terms of basis vectors, the duality is expressed by
\begin{eqnarray*}
	 \dbasis^{i_{1}} \w \dbasis^{i_{2}} \w \cdots \w \dbasis^{i_{p}}\(  \,\bvec_{j_{1}} \w \bvec_{j_{2}} \w \cdots \w \bvec_{j_{p}} \, \) &=& \delta^{i_{1}i_{2}\cdots i_{p}}_{j_{1}j_{2}\cdots j_{p}} \;=\; %
		\begin{vmatrix}
			\delta^{i_{1}}_{j_{1}}	&	\delta^{i_{1}}_{j_{2}}	& \cdots 	& \delta^{i_{1}}_{j_{p}} \\
			\delta^{i_{2}}_{j_{1}}	&	\delta^{i_{2}}_{j_{2}}	& \cdots 	& \delta^{i_{2}}_{j_{p}} \\
			\vdots					&	\vdots					& \ddots	& \vdots \\	
			\delta^{i_{p}}_{j_{1}}	&	\delta^{i_{p}}_{j_{2}}	& \cdots & \delta^{i_{p}}_{j_{p}}
		\end{vmatrix}.
\end{eqnarray*}
Thus, the action of simple multi-covectors (i.e. those obtained by exterior multiplication of single covectors) on simple multivectors can be expressed in terms of a determinant of numbers.\\

In some of our analyses we encounter terms of the form $\alpha_{p}(V_{q})$ where $\alpha_{p}$ is a $p-$covector and $V_{q}$ is a $q-$vector. For $p>q$ this may be termed a $(p-q)$-covector $\beta_{p-q}$ defined by the condition
\begin{eqnarray}\label{def_pqcovec}
	\(\,\alpha_{p}(V_{q}) \, \)(W_{p-q}) &\equiv& \alpha_{p}( V_{q} \w W_{p-q} )
\end{eqnarray}
for all $(p-q)$-vectors $W_{p-q}$. A computational scheme for the explicit calculation of $\beta_{p-q}$ will be presented below. \\

For a $\Lambda_{1}$ space with a metric, we can write the operation of taking the scalar product in terms of a bilinear function $\g$ so that for any two vector $U$ and $V$
\begin{eqnarray}
	\nonumber	U\cdot V &=& \g(U,V) \;=\; \g(V,U) \\
	\label{lam1_metric_V0}	\g(U,V) &=& 0 \qquad \forall V \qquad\Longleftrightarrow\quad U\;=\; 0.
\end{eqnarray}
Clearly for fixed $U$, $\g(U,-)$ is a linear function of $V$ and hence it is an element $\wt{U}$ say, of the dual space $\Lambda^{1}$. In terms of the covector $\wt{U}$, the scalar product may be expanded as
\begin{eqnarray*}
	\g(U,V) &=& \wt{U}(V).
\end{eqnarray*}
If we write
\begin{eqnarray*}
	U &=& U^{i}\bvec_{i} \qquadand \wt{U} \;=\; \wt{U}_{i}\dbasis^{i},
\end{eqnarray*}
then since $\wt{U}_{i}$ depends linearly on $U^{j}$, there is some matrix with elements $\g_{ij}$ such that 
\begin{eqnarray*}
	\wt{U}_{i} &=& \g_{ij}U^{i}
\end{eqnarray*}	
so that
\begin{eqnarray*}
	\g(U,V)	&=& \wt{U}(V) \;=\; \g_{ij}U^{i}V^{j}
\end{eqnarray*}
for any vector $V$. This establishes a relation between the components of a vector and the components of an associated covector in a space with metric properties. Condition (\ref{lam1_metric_V0}) ensures that the matrix $g_{ij}$ has an inverse which may be used to establish a metric in $\Lambda^{1}$. \\

Having established the spaces $\Lambda^{p}$ and $\Lambda_{p}$ based on $\real^{n}$, we now introduce the mappings that relate a space $\real^{n}$ to another $\real^{m}$. One of the most important mappings is that required to set up alternative or curvilinear coordinate in $\real^{n}$. This may be considered as a $1-1$ regular mapping $\chi$ of a set in arithmetic $n$-space into $\real^{n}$. If $\SS\in\real^{n}$ is a vector then $\SS=\chi(x)$, $x\in\real^{n}$, establishes the coordinates of $\SS$ to be $x=(x^{1},x^{2},\ldots,x^{n})$. For example, a vector $\SS=(\SS^{1},\SS^{2},\SS^{3})$ in arithmetic $3-$space can be coordinated by the $3-$tuple $(r,\theta,\phi)$ such that
\begin{eqnarray*}
	\chi^{-1}:	\real^{3}					&\longrightarrow&	\real^{3} \\
				\SS \;=\; (\SS^{1},\SS^{2},\SS^{3})	&\longmapsto&		(r,\theta,\phi) \;=\; \( \,(\chi^{-1})^{1}(\SS), (\chi^{-1})^{2}(\SS), (\chi^{-1})^{3}(\SS) \, \)
\end{eqnarray*}
with
\begin{eqnarray*}
	r		&=& \( (\SS^{1})^{2} + (\SS^{2})^{2} + (\SS^{3})^{2} \)^{\frac{1}{2}} \\[0.2cm]
	\theta 	&=& \sin^{-1}\( \frac{ \( (\SS^{1})^{2} + (\SS^{2})^{2} \)^{\frac{1}{2}} }{\( (\SS^{1})^{2} + (\SS^{2})^{2} + (\SS^{3})^{2} \)^{\frac{1}{2}}} \) \\[0.2cm]
	\phi	&=& \cos^{-1}\( \frac{\SS^{3}}{\( (\SS^{1})^{2} + (\SS^{2})^{2} + (\SS^{3})^{2} \)^{\frac{1}{2}}} \)
\end{eqnarray*}
for $(r,\theta,\phi)$ on a suitable domain.\\

\begin{figure}[!ht]
	\centering
	\includegraphics{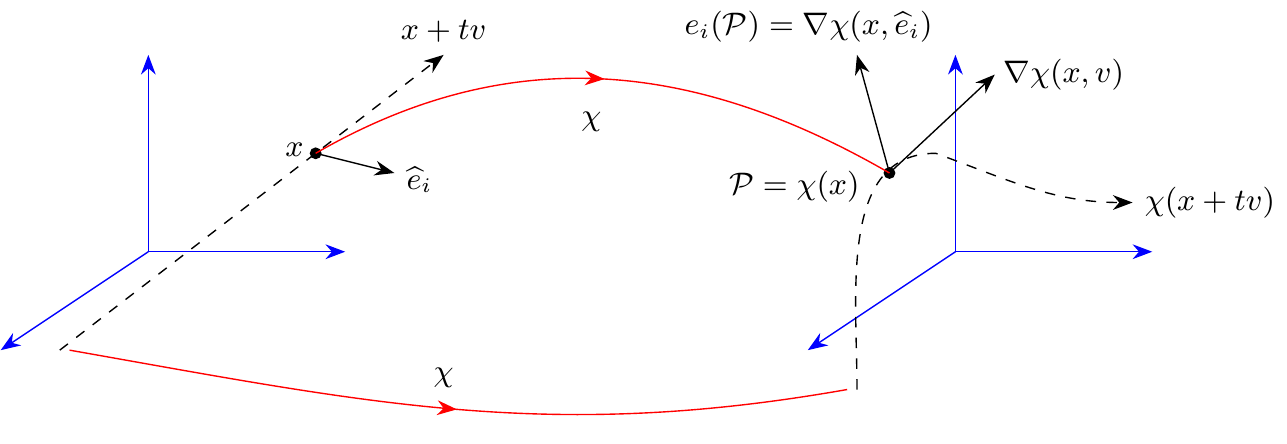}
	\caption{}\label{fig:figone}
\end{figure}

Under the mapping $\chi:\real^{n}\rightarrow\real^{n}$, the line $x+tv$ at $x\in\real^{n}$ in the direction $v$ maps into the curve $\chi(x+tv)\in\real^{n}$ parameterised by the variable $t$. The tangent vector to this curve at the point $\chi(x)$ is given by the vector
\begin{eqnarray*}
	\lim_{t\rightarrow 0}\,\frac{\chi(x+tv)-\chi(x)}{t}
\end{eqnarray*}
and is denoted by $\nabla\chi(x,v)$, the derivative of the mapping $\chi$ at $x$ along $v$. If we map, in particular, the vector $v=\wh{\bvec}_{i}$ with $\chi$, we obtain the vector
\begin{eqnarray*}
	\nonumber \bvec_{i}(\SS) \;\;\equiv\;\; \nabla\chi(x,\wh{\bvec}_{i}) &=& \lim_{t\rightarrow 0}\,\frac{\chi(x^{1},x^{2},\ldots,x^{i}+tv,\ldots x^{n})-\chi(x)}{t} \\[0.2cm]
	&=& \pdiff{\chi(x)}{x^{i}} \;\;=\;\; \pdiff{\SS}{x^{i}} \qquad (i=1,\ldots,n).
\end{eqnarray*}
as a basis vector in $\real^{n}$ (see figure~\ref{fig:figone}). Strictly speaking this vector is said to lie in the tangent space $T_{x}\real^{n}$ at $x$. \\

A common notation has evolved whereby one writes $\bvec_{i}(\SS)$ as $\pdiff{}{x^{i}}$ since the values of the co-ordinates $x^{i}$ do specify the position at which the tangent vector is evaluated. Such an entity is often referred to as a natural or co-ordinate basis vector. \\

It will be observed that working in $\real^{n}$ with general co-ordinates $x^{1},x^{2},\ldots,x^{n}$ it is necessary to specify the point $\SS$ at which one is evaluating vectors. A differential $r-$form $\omega$ in $\real^{n}$ is a function whose values $\omega(\SS)$ are $r-$ covectors, i.e. at any point $\omega(\SS)(V)$ is a number if $V$ is an $r-$vector. For any real valued smooth function $\phi:\real^{n}\rightarrow \real$ of $n$ variables, then $\nabla\phi(\SS,e)$ is a number where $\SS,e\in\Lambda_{1}$. Thus, $\nabla\phi(\SS)$ is a covector and $\nabla\phi$ a differential form. If we expand the covector $\nabla\phi(\SS)$ in a basis
\begin{eqnarray}\label{nab_phi_basis}
	\nabla\phi(\SS) &=& \gamma_{i}(\SS)\dbasis^{i}(\SS)
\end{eqnarray}
then
\begin{eqnarray*}
	\gamma_{i}(\SS) &=& \nabla\phi(\SS,e_{i}) \;\;=\;\; \pdiff{\phi(\SS)}{x^{i}}.
\end{eqnarray*}
If we chose for $\phi(\SS)$ the $j-$th component function $x^{j}(\SS)$ (i.e. the inverse of $\SS=\chi(x)$) then we see that the covector $\nabla x^{j}(\SS)$ has components $\delta^{j}_{i}$, i.e.
\begin{eqnarray*}
	\nabla x^{j}(\SS) &=& \delta^{j}_{i}\dbasis^{i} \;\;=\;\; \dbasis^{j}(\SS).
\end{eqnarray*}
In this co-ordinate basis, the covector $\dbasis^{j}(\SS)$ is usually written simply $dx^{j}$, again with the $\SS$ being implicitly located by the general co-ordinates $(x^{1},x^{2},\ldots,x^{n})$. The symbol $\nabla$ which takes the gradient of a function $\phi:\real^{n}\rightarrow\real$ (which is also referred to as a $0-$form) is written $d$ so that (\ref{nab_phi_basis}) becomes the covector
\begin{eqnarray*}
	d\phi(\SS) &=& \pdiff{\phi(\SS)}{x^{i}}\,dx^{i}.
\end{eqnarray*}
In terms of a differential form we see that the function $\nabla\phi\equiv d\phi$ can be written
\begin{eqnarray*}
	d\phi &=& \pdiff{\phi}{x^{i}}\,dx^{i}.
\end{eqnarray*}
For any vector $V$
\begin{eqnarray*}
	d\phi(V) &=& V^{i}\pdiff{\phi}{x^{i}}
\end{eqnarray*}
since we can write (\ref{def_covec_basis}) as
\begin{eqnarray*}
	dx^{i}\(\pdiff{}{x^{j}}\) &=& \delta^{i}_{j}. 
\end{eqnarray*}
In this co-ordinate basis a general differential $r-$form in $\real^{n}$ may be written
\begin{eqnarray*}
	\omega &=& \sum_{i_{1} < i_{2} < \cdots \;<\; i_{r}}^{n}  \omega_{i_{1}i_{2}\cdots i_{r}}\, dx^{i_{1}} \w dx^{i_{2}} \w \cdots \w dx^{i_{r}}
\end{eqnarray*}
where $\omega_{i_{1}i_{2}\cdots i_{r}}:\real^{n}\rightarrow\real$. At a point $\SS$ in $\real^{n}$ with co-ordinates $(x^{1},x^{2},\ldots,x^{n})$ it takes as a value the covector
\begin{eqnarray*}
	\omega(\SS) &=& \sum_{i_{1} < i_{2} < \cdots \;<\; i_{r}}^{n}  \omega_{i_{1}i_{2}\cdots i_{r}}(x^{1},x^{2},\ldots,x^{n})\, dx^{i_{1}} \w dx^{i_{2}} \w \cdots \w dx^{i_{r}}
\end{eqnarray*}
The term differential $r-$form could perhaps be replaced by the term differentiable $r-$form since all our manipulations will assume that $\omega_{i_{1}i_{2}\cdots i_{r}}(x^{1},x^{2},\ldots,x^{n})$ is a function with enough smoothness to allow evaluation of the requisite number of partial derivatives at any point. \\

The natural basis introduced here for exterior forms is useful in many calculations. It must be noted, however, that many other bases can be constructed that may be better suited for particular problems. Recall that a point in $\real^{3}$ with a Euclidean metric expressed in spherical polars $(r,\theta,\phi)$ has a triad of orthogonal non-co-ordinate tangent vectors
\begin{eqnarray*}
	(\bar{\bvec}_{r},\,\bar{\bvec}_{\theta},\,\bar{\bvec}_{\phi}) &=& \(\pdiff{}{r},\frac{1}{r}\pdiff{}{\theta},\frac{1}{r\sin(\theta)}\pdiff{}{\phi}\).
\end{eqnarray*}
A vector in this space can be expressed in either basis
\begin{eqnarray*}
	V &=& \bar{V}^{r}\bar{\bvec}_{r} + \bar{V}^{\theta}\bar{\bvec}_{\theta} + \bar{V}^{\phi}\bar{\bvec}_{\phi} \;\;=\;\; V^{r}\pdiff{}{r} + V^{\theta}\pdiff{}{\theta} + V^{\phi}\pdiff{}{\phi},
\end{eqnarray*}
so the relation between co-ordinates in the two bases is
\begin{eqnarray*}
	\bar{V}^{r} &=& V^{r},\qquad \bar{V}^{\theta} \;\;=\;\; rV^{\theta}, \qquadand \bar{V}^{\phi} \;\;=\;\; r\sin\theta\,V^{\phi}. 
\end{eqnarray*}
The co-ordinate basis is particularly well suited to describe an important linear map that takes differential $r-$forms to differential $(r+1)-$forms. This operation is called exterior differentiation and the map is also symbolised by $d$:
\begin{eqnarray*}
	d: \Lambda^{r} \longrightarrow \Lambda^{r+1}.
\end{eqnarray*}
Since the map is to be linear (alternating multi-linear to be precise) it will be defined as soon as the action on a simple $r-$form is given. If
\begin{eqnarray}\label{alp}
	\alpha &=& \sum_{i_{1} < i_{2} < \cdots \;<\; i_{r}}^{n}  \alpha_{i_{1}i_{2}\cdots i_{r}} \, dx^{i_{1}} \w dx^{i_{2}} \w \cdots \w dx^{i_{r}}
\end{eqnarray}
then
\begin{eqnarray}\label{dalp}
	d\alpha &=& \sum_{j=1}^{n}\;\sum_{i_{1} < i_{2} < \cdots \;<\; i_{r}}^{n}  \pdiff{\alpha_{i_{1}i_{2}\cdots i_{r}}}{x^{j}}\, dx^{j} \w dx^{i_{1}} \w dx^{i_{2}} \w \cdots \w dx^{i_{r}}.
\end{eqnarray}
The summation is over all $j$ from 1 to $n$ so it may be possible to simplify (\ref{dalp}) with the rules of exterior algebra. The important properties of $d$ (which are independent of any particular basis) are
\begin{eqnarray*}
	d(\alpha + \beta) &=& d\alpha + d\beta \\
	d(\alpha \w \beta) &=& d\alpha \w \beta + (-1)^{r}\,\alpha\w d\beta \\
	d(d\omega) &=& 0
\end{eqnarray*}
where $\alpha$ and $\beta$ are $r$ and $s$ forms respectively. These results may be readily verified in the co-ordinate basis using (\ref{alp}) and (\ref{dalp}). For example, from (\ref{dalp}):
\begin{eqnarray*}
	d(d\omega) 	&=& \sum_{i_{1} < \cdots \;<\; i_{r}}^{n} \sum_{i,j} \frac{\partial^{2}\omega_{i_{1}\cdots i_{r}}}{\partial x^{i}\,\partial x^{j}} \, dx^{i} \w dx^{j} \w dx^{i_{1}} \w \cdots \w dx^{i_{r}} \\[0.2cm]
				&=&	\frac{1}{2}\sum_{i_{1}  < \cdots \;<\; i_{r}}^{n} \sum_{i,j} \(\frac{\partial^{2}}{\partial x^{i}\,\partial x^{j}} - \frac{\partial^{2}}{\partial x^{j}\,\partial x^{i}}  \)\omega_{i_{1}\cdots i_{r}}\, dx^{i} \w dx^{j} \w dx^{i_{1}} \w \cdots \w dx^{i_{r}} \\[0.2cm]
				&=& 0. 
\end{eqnarray*}
This result is referred to as Poincar\'{e}'s Lemma. A form $\omega$ such that $d\omega=0$ is said to be closed. It is said to be exact if it can be expressed as $d\beta$. Every exact form is closed but in a general manifold (see below) not every closed form is exact. \\

The motivation for the definition of exterior differentiation may seem obscure at this point. Ultimately one can trace it to the generalised Stoke's theorem (to be discussed) in much the same way as the 3-dimensional curl of elementary vector analysis can be defined with respect to this theorem. \\

\begin{figure}[!ht]
	\centering
	\includegraphics{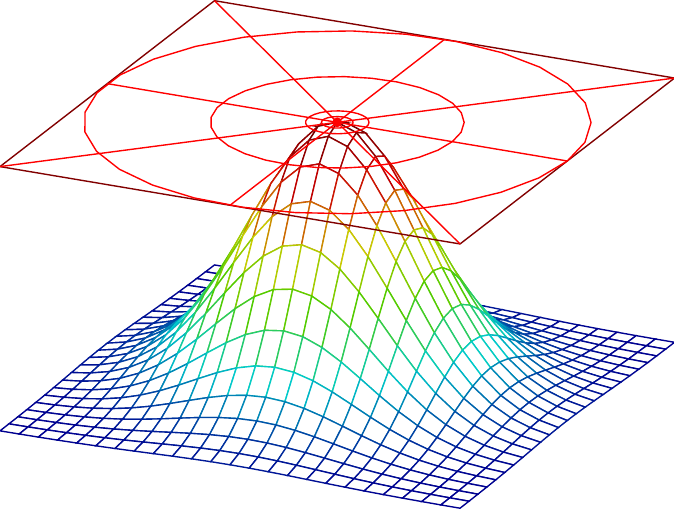}
	\caption{}\label{fig:figtwo}
\end{figure}

\begin{figure}[!ht]
	\centering
	\includegraphics{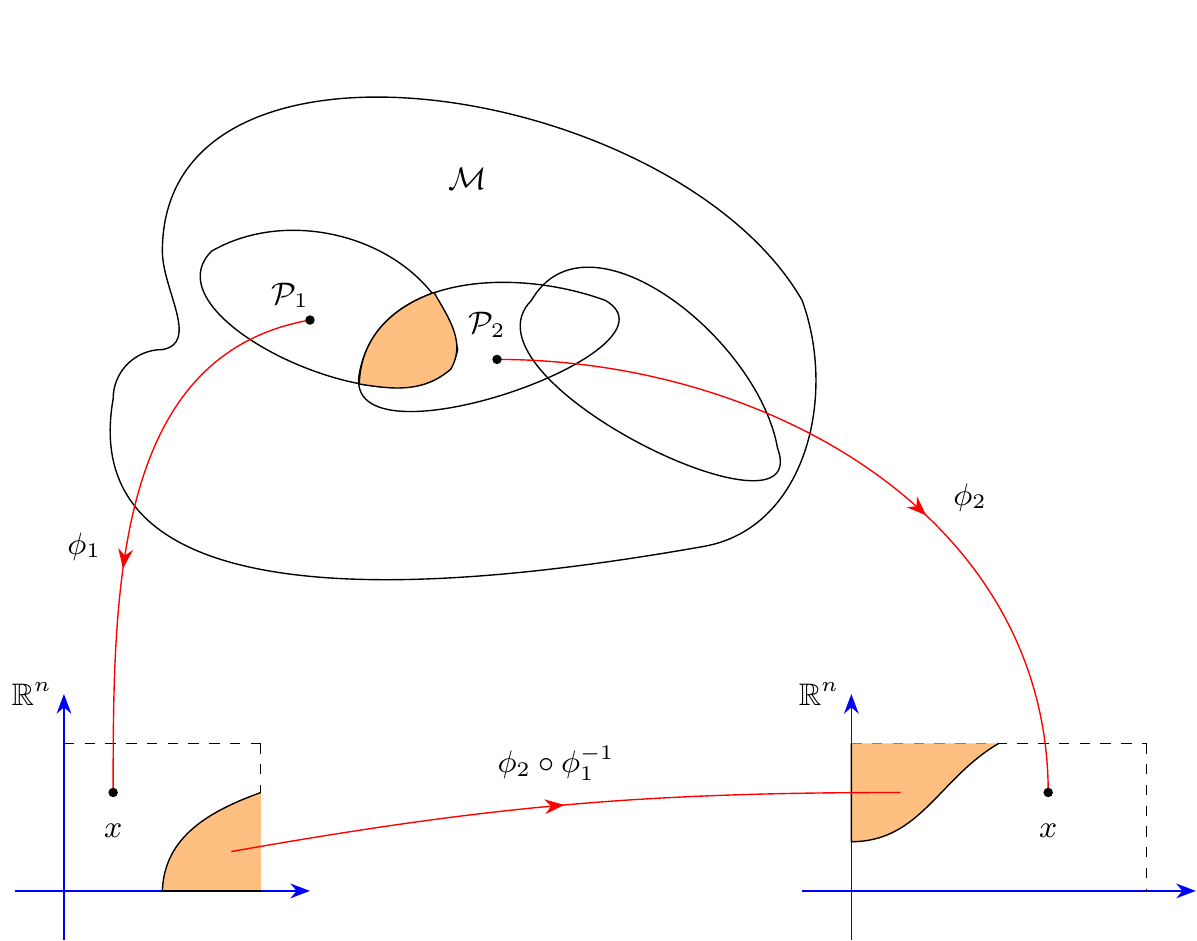}
	\caption{}\label{fig:figthree}
\end{figure}

Everything we have discussed so far has been formulated in arithmetic $n-$space. General curvilinear co-ordinates have been introduced and if necessary the space can be made Euclidean with the introduction of the Pythagorean metric. For a picture of a two-dimensional Euclidean space one may visualise a large piece of flat graph paper. It is intuitively clear that the geometry of this space differs from that on the two dimensional surface of a spherical globe. The basic idea for generalising the above formalism to smooth manifolds is to exploit one of their defining properties. At each point $p$ of an $n-$dimensional manifold there is a tangent plane $T_{p}\M$ which can be thought of as an arithmetic $n-$space, i.e. locally manifolds are like $\real^{n}$ (with or without some metric). A co-ordinate system set up in $T_{p}\M$ can be projected into some neighbourhood of $p$ on the manifold. Figure~\ref{fig:figtwo} visualises this projection in some embedding space but this is not really necessary (there is no space that we know of in which spacetime is embedded). A smooth or differentiable $n-$dimensional manifold $\M$ is a connected topological space (i.e. each point has a set of neighbourhoods) together with a set of co-ordinate maps $\phi_{1},\phi_{2},\ldots$ from $\M$ into arithmetic $n-$space. This definition is illustrated in figure~\ref{fig:figthree}. 

\begin{figure}[!ht]
	\centering
	\includegraphics{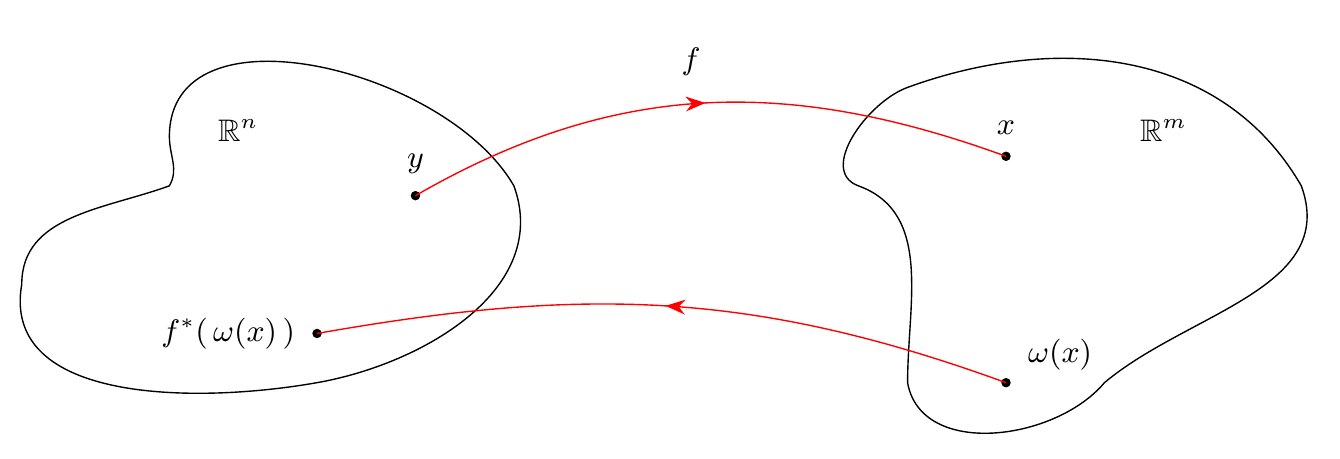}
	\caption{}\label{fig:figfour}
\end{figure}

In general more than one map $\phi$ is required to cover a manifold. If the range of two such maps $\phi_{1}$ and $\phi_{2}$ overlap then the composite function $\phi_{2}\circ \phi_{1}^{-1}$ should be a well behaved transformation. The differentiability properties of such overlap maps are in fact taken to characterise the smoothness of the manifold. The formalism of exterior forms has been established in $\real^{n}$. If we can define the effect of a mapping $\phi$ on forms and multi-vectors in $\real^{n}$ then we can establish a formalism on manifolds. We have already mapped vectors from $\real^{n}$ to $\real^{n}$ and since this space is identified with the tangent space at each point of the manifold we can write tangent vectors in terms of mappings $\{\phi_{i}\}$ belonging to the atlas that defines the manifold. If $\omega$ is a $k-$form on $\real^{m}$ and $f^{*}$ is some mapping from $\real^{m}$ to $\real^{n}$ then we can define a $k-$form $f^{*}\omega$ in $\real^{n}$ by the equation
\begin{eqnarray}\label{fomegadef}
	(f^{*}\omega)(\SS) &=& f^{*}(\,\omega(\SS)\,), \qquad f^{*}\omega \in \real^{n}. 
\end{eqnarray}
This requires a knowledge of what $f^{*}(\,\omega(\SS)\,)$ means. The least abstract way is to give the rule in a co-ordinate basis where
\begin{eqnarray*}
	\omega(\SS) &=& \sum_{i_{1} < \cdots \;<\; i_{k}}^{m}  \omega_{i_{1}\cdots i_{k}}(\SS)\,dx^{i_{1}} \w \cdots \w dx^{i_{k}}.
\end{eqnarray*}
If 
\begin{eqnarray*}
	f:\real^{n} &\longrightarrow& 	\real^{m} \\
			y	&\longmapsto&		x \;\;=\;\; f(y)
\end{eqnarray*}
then
\begin{eqnarray*}
	f^{*}(\,\omega(\SS)\,) &=& \sum_{i_{1} < \cdots \;<\; i_{k}}^{m}  \omega_{i_{1}\cdots i_{k}}(\,f(y)\,)\,\( \pdiff{f^{i_{1}}}{y^{a}}\,dy^{a}\) \w \cdots \w \( \pdiff{f^{i_{k}}}{y^{b}}\,dy^{b}\)
\end{eqnarray*}
where the indices $a,b$ are summed from 1 to $n$. The mappings are illustrated in figure~\ref{fig:figfour}. 

The rule
\begin{eqnarray*}
	f^{*}(dx^{i}) &=& \sum_{j=1}^{n}\pdiff{f^{i}}{y^{j}}\,dy^{j} \qquad i\;=\; 1,\ldots, m
\end{eqnarray*}
together with
\begin{eqnarray*}
	f^{*}(\omega_{1} + \omega_{2}) &=& f^{*}(\omega_{1}) + f^{*}(\omega_{2}) \\
	f^{*}(\omega \w \eta) &=& f^{*}(\omega) \w f^{*}(\eta)
\end{eqnarray*}
enable one to ``pull back'' forms from one space to another. What is really working here is the remarkable behaviour of the exterior derivative under a general change of co-ordinates:
\begin{eqnarray}\label{fdcom}
	f^{*}(d\omega) &=& d(\,f^{*}(\omega)\,).
\end{eqnarray}
As an explicit example of these results let us consider a three dimensional space with points co-ordinated by the functions $x$, $y$ and $z$. We examine the $2-$form
\begin{eqnarray*}
	\omega &=& x\,dy \w dz + xy^{2}\,dz \w dx + 3\,dx \w dy
\end{eqnarray*}
under the mapping
\begin{eqnarray*}
\begin{split}
	f:\real^{3} &\;\longrightarrow\; 	\real^{3} \\
		(u,v,w)	&\;\longmapsto\;		(x,y,z) \;\;=\;\; \(\,u^{2} + v,\, v^{2}+w, \, w+v\, \).   
\end{split}
\end{eqnarray*}
The $3-$form $d\omega$ is
\begin{eqnarray*}
	d\omega &=& (1+2xy)\,dx \w dy \w dz. 
\end{eqnarray*}
Thus
\begin{eqnarray*}
	f^{*}(d\omega) &=& \(\,1+2(u^{2}+v)(v^{2}+w)\,\)\, f^{*}(dx) \w  f^{*}(dy) \w  f^{*}(dz).
\end{eqnarray*}
But
\begin{eqnarray*}
	f^{*}(dx) &=& \pdiff{x}{u}\,du + \pdiff{x}{v}\,dv + \pdiff{x}{u}\,dw \;\;=\;\; 2u\,du + dv.
\end{eqnarray*}
Similarly
\begin{eqnarray*}
	f^{*}(dy) &=& 2v\,dv + dw \\
	f^{*}(dz) &=& dw + dv.
\end{eqnarray*}
So 
\begin{eqnarray*}
	f^{*}(dx) \w  f^{*}(dy) \w  f^{*}(dz) &=& \(2u\,du + dv\) \w \(2v\,dv + dw\) \w \(dw + dv \) \\[0.1cm]
		&=& 2u(2v-1)\,du \w dv \w dw
\end{eqnarray*}
giving
\begin{eqnarray}\label{fstardom}
	f^{*}(d\omega) &=& 2u(2v-1)\(\,1+2(u^{2}+v)(v^{2}+w)\,\)\, du \w dv \w dw.
\end{eqnarray}
Alternatively
\begin{eqnarray*}
\begin{split}
	f^{*}(\omega) &\;=\; (u^{2}+v)(2v\,dv + dw) \w (dw + dv ) \\
				& \; + (u^{2}+v)(v^{2}+w)^{2}\,(dx + dv) \w (2u\,du + dv) \\
				& \; + 3(2u\,du+dv) \w (2v\,dv + dw).
\end{split}
\end{eqnarray*}
Simplifying this $2-$form we obtain (\ref{fstardom}) again on exterior differentiation thus verifying (\ref{fdcom}). It should be stressed that no metric enters any of these calculations so that $(x,y,z)$ and $(u,v,w)$ need in no sense be regarded as Euclidean co-ordinates (the notion of orthogonality requires a metric for its definition). \\

A $k-$form $\omega$ on a manifold is now defined to be a function that assigns a covector $\omega(\SS)$ in the space $\Lambda^{k}$ at each point $\SS$ of the manifold. Similarly, $k-$multi-vector functions $v(\SS)$ are defined in terms of vectors from $\Lambda_{k}$ at each point on the manifold. As long as we stay at one point on the manifold a co-ordinate map can be used in conjunction with (\ref{fomegadef}) to bring the manipulations back into $\real^{n}$. However, as soon as one moves about the manifold one meets different tangent spaces each with their own co-ordinate maps back into $\real^{n}$. \\

One of the powerful properties of the general exterior differential form is that although it was defined originally in terms of comparing forms at two different point it can nevertheless be defined on a manifold independently of a particular co-ordinate system (essentially, any connections linking covectors in one tangent space to covectors in another get antisymmetrised away). The definition of $d\omega$ on the manifold relates it to the exterior derivative of the $k-$form $\omega$ in $\real^{n}$. Thus for every co-ordinate system $\phi:\M\rightarrow\real^{n}$ there is a unique $(k+1)-$form $d\omega$ such that
\begin{eqnarray*}
	\phi^{*}(d\omega) &=& d(\,\phi^{*}(\omega)\,).
\end{eqnarray*}
In many physical applications of the above formalism a manifold is the entity that is being sought from some physical principle. A field of momentum, field energy, action density etc. may be established on a manifold in some suitable space and its integral over the complete manifold required to be extremal. Thus we are led to consider integration of differential $r-$forms over manifolds \cite{ManInt}. It is convenient to think of an $m-$dimensional element of an integration domain as a cell composed of $m$ independent vectors. More precisely the $m-$domain can be constructed by joining together small domains each represented by an $m-$multi-vector. For example a closed curve may be approximated by a polygon of $1-$vectors joined end to end. A surface may be triangulated by fitting together triangles each represented by $2-$vectors $\bvec_{i}(\SS) \w \bvec_{j}(\SS)$ attached to some vertex $\SS$. A three dimensional manifold can be built up of elementary parallelepipeds represented by the $3-$vectors $\bvec_{i}(\SS) \w \bvec_{j}(\SS) \w \bvec_{k}(\SS)$. Given the elementary $r-$multi-vector $M_{i}$ at some point in the manifold and an $r-$form $\alpha$ one may evaluate the number $\alpha(\SS_{i})\,\(\,M_{i}(\SS_{i})\,\)$ and by finer and finer subdivision define the integral of an $r-$form on an $r-$manifold as
\begin{eqnarray}\label{intdef}
	\int_{\M}\alpha &=& \lim\,\sum_{i}\alpha(\SS_{i})\,\(\,M_{i}(\SS_{i})\,\). 
\end{eqnarray} 
It is interesting to observe that we have given the manifold a precise orientation at each point by associating it with a particular multi-vector. Furthermore, the degree $r$ of the form and the dimension of the manifold must match for this construction to make sense. \\

\begin{figure}[!ht]
	\centering
	\includegraphics{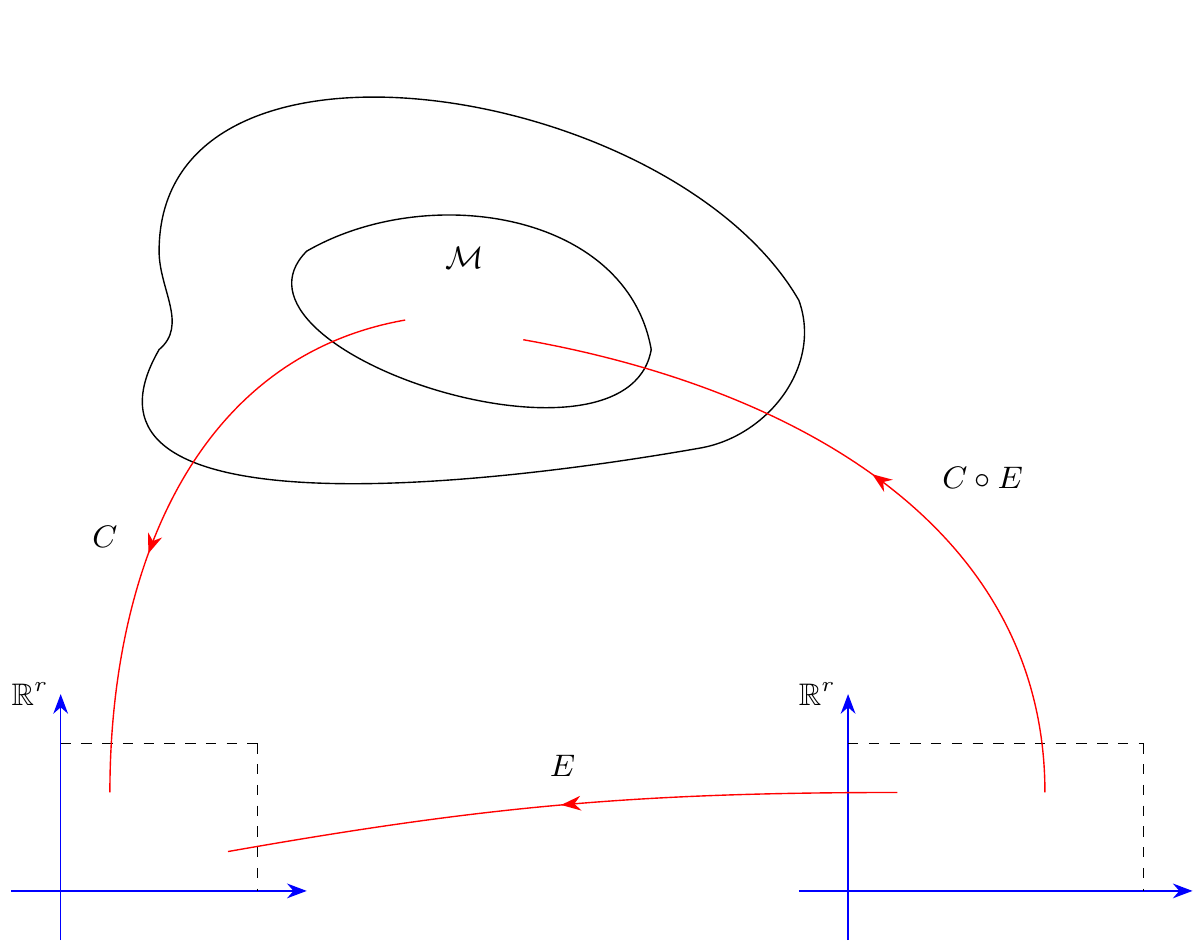}
	\caption{}\label{fig:figfive}
\end{figure}

The actual calculation of an integral such as (\ref{intdef}) is performed by using a co-ordinate mapping to bring the domain in the manifold into some standard domain in $\real^{n}$. As an example, consider the simplest one dimensional smooth manifold $\M$ embedded in two dimensional $\real^{2}$ where points are labelled with the co-ordinate function $(x,y)$. The manifold will be a curve which can be defined as the mapping $C$ of a unit interval $[0,1]\in\real$ into $\real^{2}$:
\begin{eqnarray*}
	C:[0,1] &\longrightarrow& \real^{2} \\
		t	&\longmapsto&		C(t) \;=\; (x,y) \;=\; (\,c^{1}(t),c^{2}(t)\,).
\end{eqnarray*}
If
\begin{eqnarray*}
	\omega(\SS) &=& \omega_{x}(x,y)\,dx + \omega_{y}(x,y)\,dy
\end{eqnarray*}
then
\begin{eqnarray*}
	\int_{\M}\omega &\equiv& \int_{C}\omega \;\;=\;\; \int_{[0,1]}C^{*}(\omega) \\[0.2cm]
					&=& \int_{0}^{1} \(\df \omega_{x}(c^{1}(t),c^{2}(t))\,\dot{c}^{1}(t) + \omega_{y}(C^{1}(t),c^{2}(t))\,\dot{c}^{2}(t) \)\,dt.
\end{eqnarray*}
In general if $f$ is any mapping of the region $[0,1]^{n}$ in $\real^{n}$ onto an $n-$dimensional manifold $\M$ the integral of the differential $n-$form $\omega$ is evaluated from
\begin{eqnarray*}
	\int_{\M}\alpha &=& \int_{[0,1]^{n}}f^{*}(\omega),
\end{eqnarray*}
the last integral being considered as an ordinary repeated integral. This rule can be generalised so that if $\phi:\M\rightarrow \N$ is a mapping (with a Jacobian of definite sign to preserve orientation) of the manifold $\M$ into the manifold $\N$ of the same dimension then
\begin{eqnarray*}
	\int_{\M}\phi^{*}(\omega) &=& \int_{\phi(M)}\omega \;\;\equiv\;\; \int_{\N}\omega. 
\end{eqnarray*}
This rule for the change of variables in a multiple integral can be used to show that integrals of forms over manifolds are reparameterisation invariant. This is what makes them so useful in our later applications. Reparameterisation invariance should not be confused with ordinary change of variable in an integral. As a simple example consider
\begin{eqnarray}\label{Itau}
	I_{\tau} &=& \int_{0}^{1} F\(\pdiff{x(\tau)}{\tau},\,\pdiff{y(\tau)}{\tau}\)\,d\tau,
\end{eqnarray}
an integral that depends upon the curve $(\,x(\tau),y(\tau)\,)$ in $\real^{2}$. A reparameterisation that preserves the unit interval might be
\begin{eqnarray*}
	\sigma &=& \tau^{\frac{1}{2}}. 
\end{eqnarray*}
The integral is said to be reparameterisation invariant if 
\begin{eqnarray*}
	I_{\sigma} &=& \int_{0}^{1} F\(\pdiff{x(\sigma)}{\sigma},\,\pdiff{y(\sigma)}{\sigma}\)\,d\sigma,
\end{eqnarray*}
has the same value as (\ref{Itau}). If $F(u,v)=u^{2}+v^{2}$ and the curve is $(\,x(\tau),y(\tau)\,)=(a\tau,b\tau)$ then
\begin{eqnarray*}
	I_{\tau} &=& a^{2}+b^{2} \\[0.1cm]
	I_{\sigma} &=& \frac{4}{3}\(a^{2}+b^{2}\),
\end{eqnarray*}
so (\ref{Itau}) is not reparameterisation invariant. However if $F(u,v)=(u^{2}+v^{2})^{1/2}$ then $I_{\tau}=I_{\sigma}$ and the integral is invariant. Clearly in this case any function $F$ that is homogeneous of degree one in $\dot{x}$ and $\dot{y}$ will ensure reparameterisation invariance. If we have the integral of a differential $r-$form $\omega$ over a manifold $\M$ part of which is being parameterised by the map
\begin{eqnarray*}
	C:[0,1]^{r} &\longrightarrow& \M
\end{eqnarray*}
then we can ask what will happen if we choose some other co-ordinates related to the original $\real^{r}$ by a mapping $E$
\begin{eqnarray*}
	E:[0,1]^{r} &\longrightarrow& [0,1]^{r}
\end{eqnarray*}
that preserves orientation. The new co-ordinates are given by the composition
\begin{eqnarray*}
	C\circ E:[0,1]^{r} &\longrightarrow& \M
\end{eqnarray*}
(see figure~\ref{fig:figfive}). Thus we deduce
\begin{eqnarray*}
	\int_{C\circ E}\omega &=& \int_{[0,1]^{r}} (C\circ E)^{*}(\omega) \;=\; \int_{[0,1]^{r}} E^{*}\(\,C^{*}(\omega)\,\) \;=\; \int_{E([0,1]^{r})} C^{*}(\omega) \\
		&=& \int_{[0,1]^{r}}C^{*}(\omega) \;=\; \int_{C}\omega
\end{eqnarray*}
which establishes the co-ordinate independence of this integral over the manifold. It is worthwhile spelling out in detail what lies behind these formal manipulations for a particular case of interest.\\

Suppose we consider a $2-$dimensional manifold $\M$ with point $\SS$. In some neighbourhood $\U$ of $\SS$ let the map $\phi$ co-ordinate the points by
\begin{eqnarray*}
\begin{split}
	\phi:\U &\;\longrightarrow\; \real^{2} \\
	\SS &\;\longmapsto\; \phi(\SS) \;=\; (\theta,\phi). 
\end{split}
\end{eqnarray*}
A natural basis for $2-$forms on $\M$ in this co-ordinate system is $d\theta \w d\phi$. As an illustration let us work with the particular $2-$form 
\begin{eqnarray*}
	\omega &=& \theta\,d\theta \w d\phi.
\end{eqnarray*}
Consider the reparameterisation of the manifold defined by
\begin{eqnarray*}
\begin{split}
	C:[0,1]^{2} &\;\longrightarrow\; \M \\
	(\tau,\sigma) &\;\longmapsto\; (\theta,\phi) \;=\; (\,c^{1}(\tau,\sigma),c^{2}(\tau,\sigma)\,). 
\end{split}
\end{eqnarray*}
The integral of $\omega$ over the manifold parameterised by $C$ is then
\begin{eqnarray*}
	I 	&=& \int_{C} \theta\,d\theta \w d\phi \;=\; \int_{[0,1]^{2}} C^{*}\( \theta\,d\theta \w d\phi \).
\end{eqnarray*}
Now
\begin{eqnarray*}
	C^{*}\( \theta\,d\theta \w d\phi \) &=& \theta(\sigma,\tau)\( \pdiff{\theta}{\sigma}\,d\sigma +\pdiff{\theta}{\tau}\,d\tau \) \w \( \pdiff{\phi}{\sigma}\,d\sigma +\pdiff{\phi}{\tau}\,d\tau \) \\[0.2cm]
		&=& c^{2}(\tau,\sigma)\( \pdiff{c^{2}}{\tau}\pdiff{c^{1}}{\sigma} - \pdiff{c^{2}}{\sigma}\pdiff{c^{1}}{\tau}\) \,d\sigma \w d\tau.
\end{eqnarray*}
Hence
\begin{eqnarray*}
	I 	&=& \int_{0}^{1}\int_{0}^{1} c^{2}(\tau,\sigma)\,\pdiff{(c^{2},c^{1})}{(\tau,\sigma)}\,d\sigma\,d\tau.
\end{eqnarray*}
Under the reparameterisation $[0,1]^{2}\rightarrow [0,1]^{2}$ by
\begin{eqnarray*}
\begin{split}
	\tau &\;\longrightarrow\; \tau'(\tau,\sigma) \\
	\sigma &\;\longrightarrow\; \sigma'(\tau,\sigma), 
\end{split}
\end{eqnarray*}
then
\begin{eqnarray*}
	\pdiff{(c^{2},c^{1})}{(\tau,\sigma)} &=& \pdiff{(c^{2},c^{1})}{(\tau',\sigma')}\pdiff{(\tau',\sigma')}{(\tau,\sigma)} \\[0.2cm]
	d\sigma\,d\tau &=& \pdiff{(\sigma,\tau)}{(\sigma',\tau')} \, d\sigma' \, d\tau'
\end{eqnarray*}
and hence
\begin{eqnarray*}
	I 	&=& \int_{0}^{1}\int_{0}^{1} c^{2}(\tau',\sigma')\,\pdiff{(c^{2},c^{1})}{(\tau',\sigma')}\,d\tau'\,d\sigma'
\end{eqnarray*}
which is the reparameterised form of the integral. \\

The function
\begin{eqnarray*}
	f:\real^{n} &\longrightarrow& \M
\end{eqnarray*}
that maps the unit ``cube'' $[0,1]^{n}$ is elevated to the status of a special map called a singular $n-$cube and linear combinations of such $n-$cubes with integer coefficients
\begin{eqnarray*}
	C &=& \sum_{i} n_{i}f_{i}
\end{eqnarray*}
are called $n-$chains (a $0-$chain can be considered as a map that acting on a field evaluates it at a set of points). The reason for this concept is that it allows one to calculate the boundary $\partial C$ of a chain in a well-defined manner. The straightforward combinatorics of such a calculation will not be given here, but we note that the boundary of a $1-$chain is the $0-$chain (set of points), the boundary of the square $[0,1]^{2}$ is the $1-$chain perimeter and the boundary of the cube $[0,1]^{3}$ is the set of six oriented faces constituting a $2-$chain. \\

Since the chain is a map any complicated boundary structure of a manifold can be neatly mapped into the orderly boundary structure of sets of simplices in $\real^{n}$. Of particular importance is the result that the boundary of any chain that is itself a boundary is zero:
\begin{eqnarray*}
	\partial^{2}C &=& 0.
\end{eqnarray*}
Note that the operator $\partial$ takes $n-$chains into $(n-1)-$chains compared with the operator $d$ that takes $n-$forms into $(n+1)-$forms and obeys $d^{2}=0$. \\

We now have enough apparatus (and jargon) to state the crowning theorem that relates integrals of forms on manifolds to integrals of their exterior derivative over the boundary of the manifold. It is the key formula in most of our applications and is the raison d'\^{e}tre for the definition of $d$. Given a differential $(r-1)-$form $\omega$ and an $r-$chain $C$:
\begin{eqnarray}\label{Stokes}
	\int_{C}d\omega &=& \int_{\partial C}\omega.
\end{eqnarray}
This powerful theorem reduces to many known results in manifolds of low dimension and is generally referred to as Stokes' theorem.\\
\newpage

%%%%%%%%%%%%%%%%%%%%%%%%%%%%%%%%%%%%%%%%%%%%%%%%%%%%%%%%%%%%%%%%%%%%%%%%%%%%%%
\section{The Relativistic Point Particle}\label{Sect:RelPoint}
%%%%%%%%%%%%%%%%%%%%%%%%%%%%%%%%%%%%%%%%%%%%%%%%%%%%%%%%%%%%%%%%%%%%%%%%%%%%%%
Our first application may seem rather trivial. It does however illustrate a number of points that are generalised in the subsequent applications. A free point particle has associated with it a particular worldline in spacetime. This may be regarded as a $1-$dimensional submanifold immersed in an $4-$dimensional manifold endowed with a Minkowskian metric\footnote{Throughout these notes, the matrix with components $g_{\mu\nu}$ has eigenvalues $\{1,-1,-1,-1\}$.} $g$. In the usual formulation the principle of stationary action dictates that the worldline joining two events should have an extremal ``length'' calculated with the metric of spacetime. The equation of the path can be considered as the mapping $x^{\mu}=\lambda^{\mu}(\tau)$ from some interval in $\real$ into spacetime with points labelled by $x^{\mu}$:
\begin{eqnarray*}
\begin{split}
	\mathcal{C}:	\real	&\;\longrightarrow\;	\real^{4} \\
					\tau	&\;\longmapsto\;		x^{\mu} \;=\; \lambda^{\mu}(\tau).
\end{split}
\end{eqnarray*}
If this mapping is varied then different worldlines will occur in spacetime. We seek a $1-$form in $4-$dimensions that can be integrated along some $1-$chain. The forms of interest in physical applications may depend upon $x^{\mu}$ (such as external electromagnetic fields) or on geometrical properties of the subset over which we integrate. \\

The way to formulate the problem is to allow the $1-$form to depend upon $x^{\mu}$ and the co-ordinates of the tangent vector to the worldline. The worldline itself is now regarded as a projection from some curve in a phase space with co-ordinates $x^{\mu}$ and $p^{\mu}$ where $p^{\mu}$ is proportional to the components of this tangent vector. Just as in conventional Hamiltonian dynamics $x^{\mu}$ and $p^{\mu}$ are regarded as independent co-ordinates during the variational procedure and any extremal manifold established in phase space can be projected into the configuration space with co-ordinates $x^{\mu}$. We assume that the particle is massive so that with a time-like Minkowski metric (adopted throughout these notes) the tangent vector to the worldline has positive length. Special relativity imposes the condition that once normalised the length of the vector is invariant. In order to secure Lorentz invariance we must ensure that this condition is not violated. In phase space this mean that we are working in a $7-$dimensional manifold defined by the condition
\begin{eqnarray}\label{point_con}
	p^{2} &=& m^{2}
\end{eqnarray}
where $m$ is some constant (identified with the mass of the particle). A general $1-$form in a space with co-ordinate functions $x^{\mu}$ and $p^{\mu}$ would take the form
\begin{eqnarray*}
	\omega &=& a_{\mu}\,dx^{\mu} + b_{\mu}\,dp^{\mu}
\end{eqnarray*}
in a co-ordinate basis. The tangent vector $\pdiff{}{\tau}$ at a point $\tau\in[0,1]$ can be mapped with a $1-$chain into the curve $x^{\mu}=\lambda^{\mu}(\tau)$, $p^{\mu}=\phi^{\mu}(\tau)$ in phase space. The tangent vector to this curve is then given by
\begin{eqnarray*}
	\mathcal{C}_{*}\pdiff{}{\tau} &=& \dot{\lambda}^{\mu}\,\pdiff{}{x^{\mu}} + \dot{\phi}^{\mu}\,\pdiff{}{p^{\mu}} 
\end{eqnarray*}
where $\dot{\alpha}\equiv\pdiff{\alpha}{\tau}$ for any $\alpha$. The above is written
\begin{eqnarray*}
	V &=& V_{x} + V_{p} 
\end{eqnarray*}
(N.B. $x$ and $p$ are just convenient labels on the vectors $V_{x}$ and $V_{p}$). We ensure (\ref{point_con}) by identifying\footnote{Throughout these notes, indices are raised and lowered using the appropriate metric tensor components. Thus, $\dot{\lambda}_{\mu}=g_{\mu\nu}\,\dot{\lambda}^{\nu}$.}
\begin{eqnarray}\label{phimuPoint}
	\phi^{\mu}(\tau) &=& \frac{m\dot{\lambda}^{\mu}(\tau)}{\sqrt{\dot{\lambda}^{2}\;}}
\end{eqnarray}
where $\dot{\lambda}^{2}=\dot{\lambda}^{\mu}\,\dot{\lambda}_{\mu}$. This simply states that our dynamical variables are position $x^{\mu}$ and velocity $\dot{x}^{\mu}=p^{\mu}/m$. Having chosen our variables we must select an appropriate $1-$form so that the equation of motion can be obtained by demanding that the action
\begin{eqnarray*}
	S &=& \int_{\mathcal{C}_{(1)}}\Pi
\end{eqnarray*}
be stationary under arbitrary deformations of the $1-$chain
\begin{eqnarray*}
\begin{split}
	\mathcal{C}_{(1)}:[0,1] &\;\longrightarrow\; 	\real^{7} \\
		\tau	&\;\longmapsto\;		\{ (x^{\mu}\;=\;\lambda^{\mu}(\tau), p^{\mu}\;=\; \phi^{\mu}(\tau)) \;\;|\;\; p^{2}=m^{2}\}.
\end{split}
\end{eqnarray*}

\begin{figure}[!ht]
	\centering
	\includegraphics{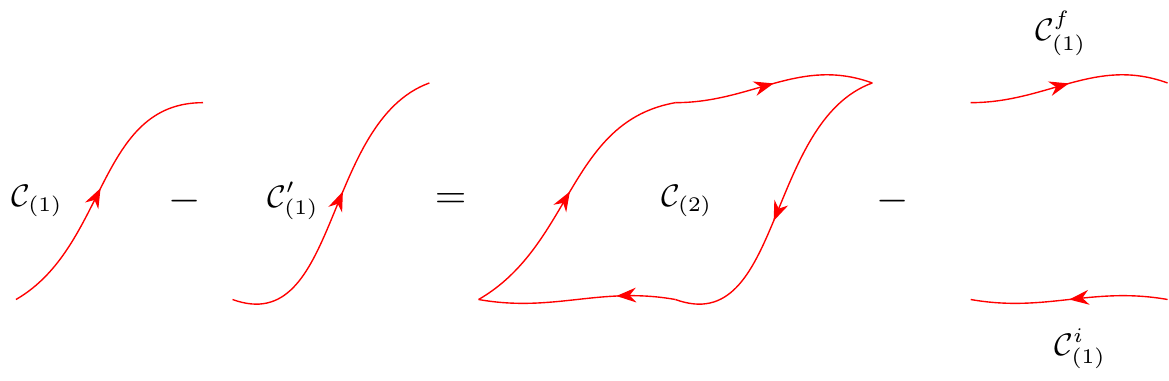}
	\caption{}\label{fig:figsix}
\end{figure}

It is convenient in the discussion to let the values of $\lambda^{\mu}(0)$ and $\lambda^{\mu}(1)$ be finite. The differential $1-$form is chosen to be
\begin{eqnarray}\label{defPiPoint}
	\Pi &\equiv& p_{\mu}\,dx^{\mu}
\end{eqnarray}
where
\begin{eqnarray*}
	p_{\mu} &=& g_{\mu\nu}\,p^{\nu}. 
\end{eqnarray*}
Let us displace the chain $\mathcal{C}_{(1)}$ in phase space to $\mathcal{C}'_{(1)}$ as illustrated in figure~\ref{fig:figsix} and calculate the difference
\begin{eqnarray*}
	\delta S &=& \int_{\mathcal{C}_{(1)}} \Pi - \int_{\mathcal{C}'_{(1)}} \Pi \;=\; \int_{\mathcal{C}_{(1)}-\mathcal{C}'_{(1)}} \Pi \;=\; \int_{\delta \mathcal{C}_{(2)}} \Pi - \(\int_{\mathcal{C}^{i}_{(1)}} \Pi  + \int_{\mathcal{C}^{f}_{(1)}} \Pi \) .
\end{eqnarray*}
The $1-$chain has been closed into the boundary $\partial \mathcal{C}_{(2)}$ of a $2-$chain by the addition of two oriented $1-$chains $\mathcal{C}^{i}_{(1)}$ and $\mathcal{C}^{f}_{(1)}$. The generalised action principle demands that
\begin{eqnarray*}
	\int_{\partial\mathcal{C}_{(2)}}\Pi &=& 0
\end{eqnarray*}
and 
\begin{eqnarray}\label{IntPiOne}
	\int_{\mathcal{C}^{i}_{(1)}}\Pi &=& \int_{\mathcal{C}^{f}_{(1)}}\Pi.
\end{eqnarray}
By Stokes' theorem (\ref{Stokes}) the first equation gives
\begin{eqnarray}\label{IntPiTwo}
	\int_{\mathcal{C}_{(2)}}d\Pi &=& 0.
\end{eqnarray}
$\mathcal{C}_{(2)}$ describes a $2-$dimensional manifold in phase space and we can regard (\ref{IntPiTwo}) as the sum
\begin{eqnarray*}
	\sum_{\SS} d\Pi(\SS)(\,V(\SS) \w \bvec(\SS)\,) &=& 0
\end{eqnarray*}
over this manifold. But
\begin{eqnarray*}
	d\Pi(\,V(\SS) \w \bvec(\SS)\,) &=& d\Pi(V) (\bvec)
\end{eqnarray*}
and since the deformation vector field $\bvec(\SS)$ is arbitrary we must have the local Euler-Lagrange equations
\begin{eqnarray*}
	d\Pi(V) &=& 0.
\end{eqnarray*}
Observe that this is the evaluation of a $2-$form $d\Pi$ with a vector so that at each point it is an equation for a covector. Let us express this equation in the co-ordinate basis $(x^{\mu},p^{\mu})$. We have
\begin{eqnarray*}
	d\Pi &=& dp_{\mu} \w dx^{\mu} \\[0.2cm]
	d\Pi(V_{x}+V_{p}) &=& dp_{\mu} \w dx^{\mu}\( \dot{\lambda}^{\mu}\,\pdiff{}{x^{\mu}} + \dot{\phi}^{\mu}\,\pdiff{}{p^{\mu}} \).
\end{eqnarray*}
But
\begin{eqnarray*}
	dp_{\mu} \w dx^{\mu}\(\pdiff{}{x^{\alpha}}\) &=& dp_{\alpha} \\[0.2cm]
	dp_{\mu} \w dx^{\mu}\(\pdiff{}{p^{\alpha}}\) &=& dp^{\mu} \w dx_{\mu}\(\pdiff{}{p^{\alpha}}\) \;=\; -dx_{\alpha}
\end{eqnarray*}
where the indices have been elevated with a constant metric $g_{\mu\nu}$ here. Thus
\begin{eqnarray*}
	d\Pi(V) &=& -\dot{\lambda}^{\alpha}\,dp_{\alpha} - \dot{\phi}^{\alpha}\, dx_{\alpha} \;=\;  - \dot{\phi}^{\alpha}\, dx_{\alpha}
\end{eqnarray*}
since 
\begin{eqnarray*}
	p^{\alpha}\,dp_{\alpha} &=& 0
\end{eqnarray*}
by (\ref{point_con}). This $1-$form must be zero so we get as equations of motion for its components
\begin{eqnarray}\label{PointEOM}
	\dot{\phi}^{\alpha} \;=\; \diff{p^{\alpha}}{\tau} &\equiv& m\diff{}{\tau}\(\frac{\dot{\lambda}^{\alpha}}{\sqrt{\dot{\lambda}^{2}\;}}\) \;=\; 0. 
\end{eqnarray}
If the parameter $\tau$ is chosen as the arc length (proper time) then $\dot{\lambda}^{2}=1$ and the world line is the path
\begin{eqnarray*}
	x^{\mu} &=& \lambda^{\mu}(\tau) \;=\; x^{\mu}(0) + \dot{x}^{\mu}(0)\,\tau.
\end{eqnarray*}
Equation (\ref{IntPiOne}) can be reduced locally to
\begin{eqnarray*}
	\Pi(\bvec_{i}) &=& \Pi(\bvec_{f})
\end{eqnarray*}
which in this case merely confirms (\ref{PointEOM}) by requiring $p^{\alpha}$ to be a constant of the motion (conservation of particle $4-$momentum). \\

A more interesting path can be obtained if the particle is electrically charged and interacts with an externally applied electromagnetic field described by a vector $4-$potential
\begin{eqnarray*}
	A &=& A_{\mu}(x)\,dx^{\mu}.
\end{eqnarray*}
This can be coupled to the particle's worldline to give a total action
\begin{eqnarray*}
	S &=& \int_{\mathcal{C}_{(1)}}\Pi + q\int_{\mathcal{C}_{(1)}} A
\end{eqnarray*}
where $q$ is the electric charge of the particle. Note incidentally that if the $1-$form $A$ is modified by the addition of the particular $1-$form $d\phi$ where $\phi$ is any $0-$form
\begin{eqnarray*}
	\int_{\mathcal{C}_{(1)}} (A+d\phi) &=& \int_{\mathcal{C}_{(1)}} A + \int_{\partial\mathcal{C}_{(1)}} \phi \;=\; \int_{\mathcal{C}_{(1)}} A + \phi(1) - \phi(0) 
\end{eqnarray*}
so that if $\phi$ vanishes at the extremities of the worldline then the interaction is invariant under this transformation. The principle of stationary action now yields
\begin{eqnarray}\label{dPiA}
	d(\Pi + qA)(V) &=& 0
\end{eqnarray}
and the conservation equation
\begin{eqnarray*}
	(\Pi+qA)(\bvec_{i}) &=& (\Pi+qA)(\bvec_{f}).
\end{eqnarray*}
Since $A$ depends only on $x$ we find in a co-ordinate basis
\begin{eqnarray*}
	dA(V) 	&=& dA_{\mu} \w dx^{\mu}\( \dot{\lambda}^{\nu}\,\pdiff{}{x^{\nu}} + \dot{\phi}^{\nu}\,\pdiff{}{p^{\nu}} \) \\[0.2cm]
			&=& \pdiff{A_{\mu}}{x^{\beta}}\,dx^{\beta} \w dx^{\mu} \( \dot{\lambda}^{\nu}\,\pdiff{}{x^{\nu}} \) \\[0.2cm]
			&=& \pdiff{A_{\mu}}{x^{\beta}}\,\dot{\lambda}^{\nu}\,\(dx^{\beta} \, \delta^{\mu}_{\nu} -  dx^{\mu}\,\delta^{\beta}_{\nu} \) \\[0.2cm]
						&=& \( \pdiff{A_{\mu}}{x^{\beta}} - \pdiff{A_{\beta}}{x^{\mu}}\)\dot{\lambda}^{\mu}\,dx^{\beta}.
\end{eqnarray*}
In this case we obtain the equation of motion for the components
\begin{eqnarray}\label{PointEOM}
	m\diff{}{\tau}\(\frac{\dot{\lambda}^{\beta}}{\sqrt{\dot{\lambda}^{2}\;}}\) \;=\; q\,F_{\beta\mu}\,\dot{\lambda}^{\mu}. 
\end{eqnarray}
Thus it is the $4-$dimensional curl ($F=dA$) that occurs in the equation of motion. Note that as a result of $d^{2}=0$ we have
\begin{eqnarray*}
	dF &=& 0
\end{eqnarray*}
which is one of Maxwell's equations for the external field. \\

Gravitational interactions with the particle can be accommodated with equal ease. We simply evaluate (\ref{dPiA}) in an arbitrary $x^{\mu}$ co-ordinate system and recall that
\begin{eqnarray}
	\nonumber \Pi &=& g_{\mu\beta}(x)\,p^{\beta}\,dx^{\mu} \\
	\label{PointGpp} g_{\mu\alpha}(x)\,p^{\mu}\,p^{\alpha} &=& m^{2}.
\end{eqnarray}
The exterior derivative of $\Pi$ now becomes\footnote{We adopt the common notation $\displaystyle g_{\mu\nu,\alpha}=\pdiff{g_{\mu\nu}}{x^{\alpha}}$.}
\begin{eqnarray*}
	d\Pi &=& p^{\nu}\,g_{\mu\nu,\alpha}\,dx^{\alpha} \w dx^{\mu} + g_{\mu\nu}\,dp^{\nu} \w dx^{\mu}
\end{eqnarray*}
so that
\begin{eqnarray}\label{PointdPiVA}
	d\Pi(V) &=& \dot{\lambda}^{\alpha}\,p^{\nu}\,\(g_{\alpha\nu,\beta} - g_{\beta\nu,\alpha}\)\,dx^{\beta} + \dot{\lambda}^{\alpha}\,g_{\alpha\nu}\,dp^{\nu} - g_{\mu\nu}\,\dot{p}^{\nu}\, dx^{\mu}
\end{eqnarray}
The penultimate term is no longer zero but
\begin{eqnarray*}
	\dot{\lambda}^{\alpha}\,g_{\alpha\nu}\,dp^{\nu} &=& -\frac{\sqrt{\dot{\lambda}^{2}\;}}{2m}\, g_{\alpha\mu,\beta}\,p^{\alpha}\,p^{\mu}\,dx^{\beta}
\end{eqnarray*}
from (\ref{PointGpp}). Substituting into (\ref{PointdPiVA}) gives immediately
\begin{eqnarray*}
	m\diff{}{\tau}\(\frac{\dot{\lambda}^{\beta}}{\sqrt{\dot{\lambda}^{2}\;}}\) + \frac{m}{\sqrt{\dot{\lambda}^{2}\;}}\,\Gamma^{\beta}_{\alpha\nu}\,\dot{\lambda}^{\alpha}\,\dot{\lambda}^{\nu} \;=\; q\,F^{\beta}_{\;\mu}\,\dot{\lambda}^{\mu}
\end{eqnarray*}
as the co-ordinate representation of (\ref{dPiA}) with the particle in a curved spacetime (in which case the Christoffel symbols $\Gamma^{\beta}_{\alpha\nu}(x)$ derived from the derivatives of metric components cannot all be set to zero by a change of co-ordinates). The square roots can be replaced by unity if the worldline is parameterised by its arc length.\\
\newpage

%%%%%%%%%%%%%%%%%%%%%%%%%%%%%%%%%%%%%%%%%%%%%%%%%%%%%%%%%%%%%%%%%%%%%%%%%%%%%%
\section{The Relativistic String}\label{Sect:RelString}
%%%%%%%%%%%%%%%%%%%%%%%%%%%%%%%%%%%%%%%%%%%%%%%%%%%%%%%%%%%%%%%%%%%%%%%%%%%%%%
We develop the theory of the classical relativistic string in close analogy with the free relativistic particle of the previous section. The interest in extended classical systems arises for a number of reasons. It is thought that they might arise as localised solutions of certain underlying gauge field theories \cite{Gauge} and the relevance of the latter to high energy physics needs to further discussion. Having spatial extension such systems contain degrees of freedom that can be dynamically excited. Translated into relativistic terms they may be expected to sustain a mass spectrum. The interest in the relativistic string is greatly increased by the observation that its (mass)${}^{2}$ excitation spectrum varies linearly with angular momentum in good accord with observation. The string is conceived classically as approximating a bounded $1-$dimensional space-like manifold immersed in spacetime. A novel feature is the possible existence of different spatial topologies for the string. Let us concentrate at first on the closed string which may be pictured relativistically as a tube-like $2-$dimensional manifold $M_{x}$ in spacetime. We therefore seek a $2-$form $\Pi$ which will render the action
\begin{eqnarray*}
	S &=& \frac{\hbar}{\alpha'}\int_{\mathcal{A}}\Pi
\end{eqnarray*}
stationary on some $2-$dimensional phase space chain $\mathcal{A}$. Let us describe the manifold $M_{x}$ by the chain
\begin{eqnarray*}
\begin{split}
	\mathcal{A}_{x}:	\real^{2}	&\;\longrightarrow\;	\real^{4} \\
					(\sigma,\tau)	&\;\longmapsto\;		x^{\mu} \;=\; \lambda^{\mu}(\sigma,\tau), \qquad \sigma\in[0,1]
\end{split}
\end{eqnarray*}
where for the closed string $\lambda^{\mu}(0,\tau)=\lambda^{\mu}(1,\tau)$. The $2-$vector $\pdiff{}{\tau}\w\pdiff{}{\sigma}$ mapped into spacetime becomes
\begin{eqnarray*}
	A_{x*}\(\pdiff{}{\tau}\w\pdiff{}{\sigma}\) &=& \frac{1}{2}\dot{\lambda}^{[\nu}\,\lambda'^{\mu]}\,\pdiff{}{x^{\mu}}\w\pdiff{}{x^{\nu}}
\end{eqnarray*}
where dot and prime refer to $\tau$ and $\sigma$ partial differentiation respectively. Since the squared norm of this $2-$vector is
\begin{eqnarray*}
	\frac{1}{4} \dot{\lambda}_{[\mu}\,\lambda'_{\nu]} \,\dot{\lambda}^{[\alpha}\,\lambda'^{\beta]} \( dx^{\mu} \w dx^{\nu} \)\( \pdiff{}{x^{\alpha}}\w\pdiff{}{x^{\beta}}\) &=& \dot{\lambda}^{2}\lambda'^{2} - (\dot{\lambda}\cdot\lambda')^{2}
\end{eqnarray*}
we can define a unit norm momentum $2-$form
\begin{eqnarray}\label{defStringPi}
	\Pi &=& \pi_{\mu\nu}\,dx^{\mu} \w dx^{\nu}
\end{eqnarray}
where
\begin{eqnarray}\label{defStringPicomp}
	\pi_{\mu\nu} &=& \frac{\dot{\lambda}_{[\mu}\,\lambda'_{\nu]}}{\sqrt{(\dot{\lambda}\cdot\lambda')^{2} - \dot{\lambda}^{2}\,\lambda'^{2}\;}} \;=\; -\pi_{\nu\mu}. 
\end{eqnarray}
Equation (\ref{defStringPi}) and (\ref{defStringPicomp}) may be compared with (\ref{defPiPoint}) and (\ref{phimuPoint}) for the point particle. We now define our phase space for the system to have co-ordinates $x^{\mu},\pi^{\mu\nu}$, i.e. it is the $2-$manifold defined by the chain
\begin{eqnarray*}
\begin{split}
	\mathcal{C}_{(2)}: \real^{2} 	&\;\longrightarrow\;	\real^{9} \\
					(\sigma,\tau)	&\;\longmapsto\;		\{x^{\mu}\;=\;\lambda^{\mu}(\sigma,\tau), \pi^{\mu\nu}\;=\; \phi^{\mu\nu}(\sigma,\tau) \;\;|\;\; \pi^{\mu\nu}\pi_{\mu\nu}\;=\; 1\}.
\end{split}
\end{eqnarray*}
It may be observed that the square root in (\ref{defStringPicomp}) has been taken assuming that $\dot{\lambda}^{\mu}$ and $\lambda'^{\mu}$ are the components of time-like and space-like vectors respectively. Next introduce in phase space the $1-$vectors
\begin{eqnarray*}
	V_{\sigma}	&=& \lambda'^{\mu}\,\pdiff{}{x^{\mu}} + \phi'^{\mu\nu}\pdiff{}{\pi^{\mu\nu}} \;\equiv\; V_{\sigma x} + V_{\sigma\pi} \\[0.3cm]
	V_{\tau}	&=& \dot{\lambda}^{\mu}\,\pdiff{}{x^{\mu}} + \dot{\phi}^{\mu\nu}\pdiff{}{\pi^{\mu\nu}} \;\equiv\; V_{\tau x} + V_{\tau\pi},
\end{eqnarray*}
so that we have the $2-$vector fields
\begin{eqnarray}
	\nonumber Y_{x}	&=& V_{\sigma x} \w V_{\tau x} \\[0.2cm]
	\label{Yvec} Y		&=& V_{\sigma} \w V_{\tau}.
\end{eqnarray}
The normalisation condition analogous to (\ref{point_con}) now reads
\begin{eqnarray*}
	\Pi(Y_{x}) &=& 1.
\end{eqnarray*}
If we work with a Minkowski flat metric with constant components $g_{\mu\nu}$ then from this equation
\begin{eqnarray*}
	d\Pi(Y_{x})	&\equiv&	\frac{1}{2}\,d(\pi^{\alpha\beta})\,\pi_{\alpha\beta} \;=\; 0.
\end{eqnarray*}

\begin{figure}[!ht]
	\centering
	\includegraphics{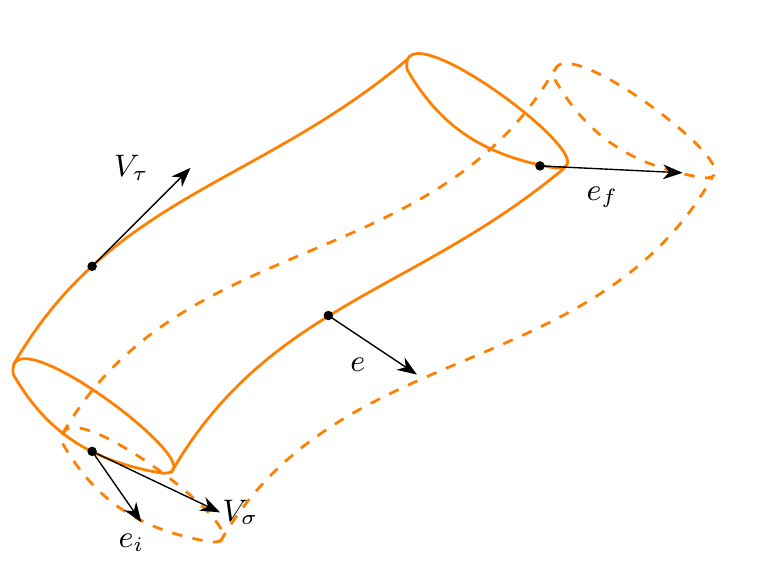}
	\caption{}\label{fig:figseven}
\end{figure}

We have now set up the problem ready for variations of $\mathcal{A}$ in phase space. We displace the $2-$manifold with a new chain $\mathcal{A}'$ and obtain (see figure~\ref{fig:figseven})
\begin{eqnarray*}
	\nonumber \delta A 	&=& \int_{\mathcal{A}-\mathcal{A}'}\Pi + \int_{V_{\sigma}\w \bvec_{i}}\Pi - \int_{V_{\sigma}\w \bvec_{f}}\Pi \\
						&=& \int_{\partial(Y\w\bvec)}\Pi + \int_{V_{\sigma}\w \bvec_{i}}\Pi - \int_{V_{\sigma}\w \bvec_{f}}\Pi
\end{eqnarray*}
where we have ``capped'' the displaced ``volume'' so that $\mathcal{A}-\mathcal{A}'$ is the boundary of some closed $3-$chain $Y\w\bvec$. The variational principle gives with the aid of Stokes' theorem
\begin{eqnarray*}
	d\Pi(Y \w \bvec) &=& 0 \\[0.2cm]
	\label{StringIntV} \int_{V_{\sigma}\w \bvec_{i}}\Pi &=& \int_{V_{\sigma}\w \bvec_{f}}\Pi.
\end{eqnarray*}
Since $\bvec$ is arbitrary, this implies the equation
\begin{eqnarray}
	\label{dPiY} d\Pi(Y) &=& 0 \\[0.2cm]
	\label{StringBC} \int \Pi(V_{\sigma x})(\bvec_{i}) &=& \int \Pi(V_{\sigma x})(\bvec_{f}).
\end{eqnarray}
We have been able to replace $V_{\sigma}$ by $V_{\sigma x}$ in (\ref{StringIntV}) since $\Pi$ is independent of $d\pi^{\mu\nu}$, i.e. $\Pi(V_{\sigma\pi})=0$. These equations may be regarded as the most general reparameterisation invariant statements about the motion of a free string system in spacetime. To obtain their co-ordinate representations is a straightforward matter of computation. Each equation is an expression of a covector field and repeated application of the rule (\ref{def_pqcovec}) can be used in this reduction to co-ordinates. For this purpose we give as illustration some results of exterior algebra that may be used:
\begin{eqnarray*}
	(\dbasis^{\mu} \w \dbasis^{\nu})(\bvec_{\alpha})	&=& \dbasis^{\mu}\,\delta_{\alpha}^{\;\nu} - \dbasis^{\nu}\,\delta_{\alpha}^{\;\mu} \\[0.2cm]
	(\dbasis^{\mu} \w \dbasis^{\nu})(\bvec_{\alpha} \w \bvec_{\beta})	&=& \delta_{\beta}^{\;\mu}\,\delta_{\alpha}^{\;\nu} - \delta_{\beta}^{\;\nu}\,\delta_{\alpha}^{\;\mu} \\[0.2cm]
	(\dbasis^{\mu} \w \dbasis^{\nu} \w \dbasis^{\rho})(\bvec_{\alpha}) &=& \dbasis^{\mu} \w \dbasis^{\nu}\,\delta_{\alpha}^{\;\rho} - \dbasis^{\mu} \w \dbasis^{\rho}\,\delta_{\alpha}^{\;\nu} + \dbasis^{\nu} \w \dbasis^{\rho}\,\delta_{\alpha}^{\;\mu}.
\end{eqnarray*}
Inserting (\ref{defStringPi}) and (\ref{Yvec}) into (\ref{dPiY}) one obtains as covariant equations for the string in flat spacetime
\begin{eqnarray}\label{StringCovEq}
	\pdiff{}{\sigma}\( \frac{\dot{\lambda}^{\nu}\,\dot{\lambda}_{[\mu}\,\lambda'_{\nu]}\,}{\Delta} \) + \pdiff{}{\tau}\( \frac{\lambda'^{\nu}\,\dot{\lambda}_{[\mu}\,\lambda'_{\nu]}\,}{\Delta} \) &=& 0
\end{eqnarray}
where 
\begin{eqnarray*}
	\Delta &=& \sqrt{ (\dot{\lambda}\cdot\lambda')^{2} - \dot{\lambda}^{2}\,\lambda'^{2} \;}.
\end{eqnarray*}
It is convenient to define the determinants
\begin{eqnarray*}
	p_{\tau}^{\,\mu}(\sigma,\tau) 	&=& \frac{1}{\Delta}	\begin{vmatrix}
																\dot{\lambda}^{\mu} & \lambda'^{\mu} \\
																\dot{\lambda}\cdot\lambda' & \lambda'^{2}
															\end{vmatrix} \\[0.3cm]
	p_{\sigma}^{\,\mu}(\sigma,\tau) 	&=& \frac{1}{\Delta}	\begin{vmatrix}
																\dot{\lambda}^{2} & \dot{\lambda}\cdot\lambda' \\
																\dot{\lambda}^{\mu} & \lambda'^{\mu}
															\end{vmatrix}
\end{eqnarray*}
and write (\ref{StringCovEq}) as the divergence condition
\begin{eqnarray*}
	\pdiff{}{\sigma}\,p_{\sigma}^{\,\mu} + \pdiff{}{\tau}\,p_{\tau}^{\,\mu} &=& 0.
\end{eqnarray*}
Once we have decided on a time co-ordinate $\tau$ we can use the vector $\bvec_{i}$ (or $\bvec_{f}$) to displace the chain $V_{\sigma x}$ in the time direction. The boundary condition (\ref{StringBC}) now says that the $1-$forms $\Pi(V_{\sigma})$ integrated over the initial and final boundary $1-$chains are the same or in co-ordinate language
\begin{eqnarray*}
	p_{\tau}^{\,\mu} &=& \int_{0} p_{\sigma}^{\,\mu}(\sigma,\tau)\,d\sigma
\end{eqnarray*}
is a constant of the motion. \\

Apart from their manifest Lorentz invariance the 4 equations (\ref{StringCovEq}) are not particularly useful. Because of the general $\sigma,\tau$ reparameterisation invariance only 2 of these equations are independent. The existence of identities follows as soon as we look at the norms of the $1-$forms
\begin{eqnarray*}
	p_{\tau} &=& \Pi(V_{\sigma x}) \\[0.1cm]
	p_{\sigma} &=& \Pi(V_{\tau x}).
\end{eqnarray*}
These particular forms play an important role in the development of the theory of strings. Once $\tau$ is chosen as an evolution parameter then $p_{\tau}$ becomes a conserved momentum $1-$form and the identities involving $p_{\tau}$ become primary constraints in a Hamiltonian formulation. $p_{\sigma}$ appears in the boundary condition for an open string to be discussed below. The co-ordinate free calculation of the identities proceeds as follows:
\begin{eqnarray*}
	p_{\tau} &=& \frac{(\wt{V}_{\tau x} \w \wt{V}_{\sigma x})(V_{\sigma x})}{\sqrt{(\wt{V}_{\tau x} \w \wt{V}_{\sigma x})(V_{\tau x} \w V_{\sigma x})\;}} \\[0.3cm]
	p_{\tau}(\wt{p}_{\tau}) &=& \frac{\[(\wt{V}_{\tau x} \w \wt{V}_{\sigma x})(V_{\sigma x})\]\,\[(V_{\tau x} \w V_{\sigma x})(\wt{V}_{\sigma x})\]}{(\wt{V}_{\tau x} \w \wt{V}_{\sigma x})(V_{\tau x} \w V_{\sigma x})} .
\end{eqnarray*}
Using (\ref{def_pqcovec}) we rewrite the numerator
\begin{eqnarray*}
	\nonumber 	&& (\wt{V}_{\tau x} \w \wt{V}_{\sigma x})\( \,V_{\sigma x} \w \(\,(V_{\tau x} \w V_{\sigma x})(\wt{V}_{\sigma x})\,\) \,\) \\[0.2cm]
	\nonumber	&& \hspace{2cm} \;=\; -(\wt{V}_{\tau x} \w \wt{V}_{\sigma x})\( \, \(\, (V_{\tau x} \w V_{\sigma x})(\wt{V}_{\sigma x}) \,\) \w V_{\sigma x}\) \\[0.2cm]
	\nonumber	&& \hspace{2cm} \;=\; -(\wt{V}_{\tau x} \w \wt{V}_{\sigma x})\( \, V_{\tau x} \(\, V_{\sigma x} \(\,\wt{V}_{\sigma x}\,\) \,\) -\(\, V_{\tau x}\(\,\wt{V}_{\sigma x}\,\)\,\)\,V_{\sigma x}   \) \w V_{\sigma x} \\[0.2cm]
				&& \hspace{2cm} \;=\; -(\wt{V}_{\tau x} \w \wt{V}_{\sigma x})\(\,V_{\tau x} \w V_{\sigma x}\,\)\(\,V_{\sigma x}(\wt{V}_{\sigma x})\,\)
\end{eqnarray*}
since $V_{\sigma x} \w V_{\sigma x}=0$. Thus
\begin{eqnarray*}
	p_{\tau}(\wt{p}_{\tau}) &=& -V_{\sigma x}(\wt{V}_{\sigma x}). 
\end{eqnarray*}
In a similar manner we find
\begin{eqnarray}\label{Stringpsigpsig}
	p_{\sigma}(\wt{p}_{\sigma}) &=& V_{\tau x}(\wt{V}_{\tau x}). 
\end{eqnarray}
Furthermore
\begin{eqnarray*}
	p_{\tau}(\wt{V}_{\sigma x}) &=& \Pi(V_{\sigma x})(\wt{V}_{\sigma x}) \;=\; \Pi(V_{\sigma x} \w \wt{V}_{\sigma x}) \;=\; 0 \\[0.1cm]
	p_{\sigma}(\wt{V}_{\tau x}) &=& \Pi(V_{\tau x})(\wt{V}_{\tau x}) \;=\; \Pi(V_{\tau x} \w \wt{V}_{\tau x}) \;=\; 0 
\end{eqnarray*}

\quad \\
\begin{figure}[!ht]
	\centering
	\includegraphics{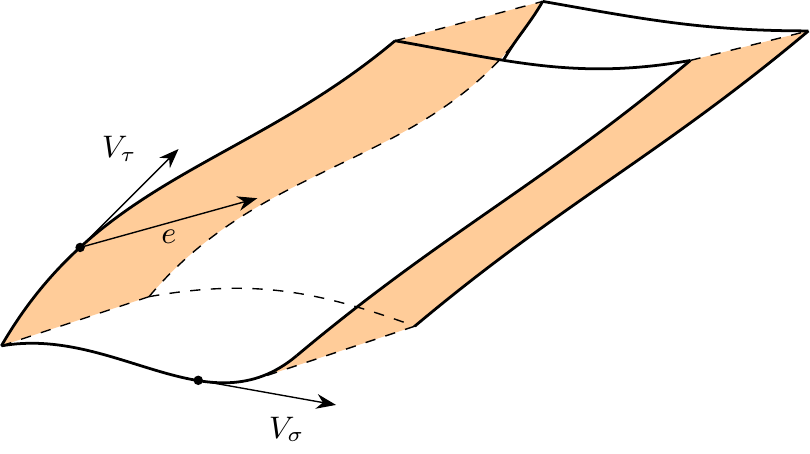}
	\caption{}\label{fig:figeight}
\end{figure}

The calculations for an open string proceed in an analogous manner expect that in the variational procedure we must add two $2-$dimensional ``panels'' in phase space (see figure~\ref{fig:figeight}) in order to render $\mathcal{A}'-\mathcal{A}$ the boundary of a closed manifold. The contributions to the action from these integrals must separately vanish so we obtain the boundary conditions
\begin{eqnarray*}
	\int_{V_{\tau}\w \bvec} \Pi &=& 0
\end{eqnarray*}
at each extremity. Since $\bvec$ is arbitrary this implies
\begin{eqnarray*}
	\Pi(V_{\tau x}) &=& 0
\end{eqnarray*}
or 
\begin{eqnarray*}
	p_{\sigma} &=& 0
\end{eqnarray*}
along each boundary in spacetime. From (\ref{Stringpsigpsig}) we see that his means that the tangent vectors to each boundary worldline of the open string are null, i.e. the string ends must move with the velocity of light. \\

We remarked earlier that the covariant equations of motion are of limited utility. String theory has progressed by utilising a remarkable geometric feature of $2-$dimensional manifolds. It is possible to find some co-ordinates, $u(\SS),v(\SS)$ say, such that the metric can always be reduced locally to the form
\begin{eqnarray*}
	ds^{2} &=& \lambda(u,v)\,\(\,du^{2} - dv^{2}\,\)
\end{eqnarray*}
and the manifold is said to be conformally flat. When such co-ordinates are established on the extremal string manifold the equations (\ref{StringCovEq}) simplify considerably. However, the real significance of the gauge freedom associated with the reparameterisation invariance is that only two field degrees of freedom determined by the covariant equations are independent. The other two (in $4-$dimensional spacetime) effectively establish a co-ordinate mesh on the $2-$manifold. Such time-like and longitudinal degrees of freedom have analogues in the vector potential $A$ for the electromagnetic field. The isolation of the independent degrees of freedom (which are those most naturally quantised in the canonical approach) can only be accomplished when the $2-$dimensional string manifold is unambiguously co-ordinated in spacetime. From our earlier discussion on defining manifolds we expect this to be a delicate job in general (see the appendix of \cite{GaugeMem} for some subtleties involved in fixing a gauge for the closed string). Partly in order to prepare the ground for the final section and partly to motivate the choice that is usually made for the string, we shall approach the problem from a rather general viewpoint.\\

Let us suppose that flat spacetime has general curvilinear co-ordinates $x^{\mu}$ with a metric $g_{\mu\nu}(x)$. The $3-$dimensional manifold given by the equation $x^{3}=0$ will in general intersect the $3-$dimensional manifold $x^{2}=0$ in a $2-$dimensional sheet. Suppose that our co-ordinates are oriented so that the string manifold has the parametric representation
\begin{eqnarray*}
	x^{\mu} &=& (x^{0},x^{1},x^{2},x^{3}) \;=\; (\tau,\sigma,0,0).
\end{eqnarray*}
If we knew how the $x^{\mu}$ were related to some global orthogonal co-ordinates, $y^{\mu}$ say, in some fixed reference frame then for a specific range of $\sigma$ and $\tau$ the surface would be defined. If the metric has constant components $\eta_{\alpha\beta}$ in the $\{y^{\mu}\}$ co-ordinate system then
\begin{eqnarray*}
	g_{\mu\nu} &=& \eta_{\alpha\beta}\,\pdiff{Y^{\alpha}}{x^{\mu}}\,\pdiff{Y^{\beta}}{x^{\nu}}
\end{eqnarray*}
where
\begin{eqnarray*}
	Y: \real^{4} &\longrightarrow& \real^{4} \\
		x^{\mu} &\longmapsto& y^{\mu} \;=\; Y^{\mu}(x).
\end{eqnarray*}
Whatever this gauge choice we can evaluate (\ref{dPiY}) in the metric $g_{\mu\nu}$. This in itself will reduce us from 4 equations to 2 for the metric coefficients on the manifold $x^{2}=x^{3}=0$. To this end let us introduce explicitly the co-ordinate basis and write
\begin{eqnarray*}
	\Pi &=& \pi_{A}\,dx^{0} \w dx^{1} + \pi_{B}\,dx^{0} \w dx^{2} + \pi_{C}\,dx^{0} \w dx^{3} + \pi_{D}\,dx^{1} \w dx^{2} \\
		&& \qquad + \pi_{E}\,dx^{1} \w dx^{3} + \pi_{F}\,dx^{2} \w dx^{3}.
\end{eqnarray*}
Then
\begin{eqnarray*}
	\pi_{A} &=& \frac{1}{\Delta}	\begin{vmatrix}
										\dot{x}_{0} & \dot{x}_{1} \\
										x'_{0} & x'_{1}
									\end{vmatrix}  \;=\;  \frac{1}{\Delta}	\begin{vmatrix}
																				g_{00} & g_{10} \\
																				g_{01} & g_{11}
																			\end{vmatrix}  \;=\; \Delta
\end{eqnarray*}
where 
\begin{eqnarray*}
	\Delta^{2} &=& 	\begin{vmatrix}
						\dot{x}^{\mu}x_{\mu} & \dot{x}^{\mu}x'_{\mu}  \\
						\dot{x}^{\mu}x'_{\mu} &  x'^{\mu}x'_{\mu} 
					\end{vmatrix}  \;=\;  	\begin{vmatrix}
												g_{00} & g_{01} \\
												g_{10} & g_{11}
											\end{vmatrix}.
\end{eqnarray*}
Similarly
\begin{eqnarray*}
	\pi_{B} &=&	\frac{1}{\Delta}	\begin{vmatrix}
										g_{00} & g_{20} \\
										g_{01} & g_{21}
									\end{vmatrix}, \qquad
	\pi_{C} \;=\; \frac{1}{\Delta}	\begin{vmatrix}
										g_{00} & g_{30} \\
										g_{01} & g_{31}
									\end{vmatrix}, \qquad
	\pi_{D} \;=\; \frac{1}{\Delta}	\begin{vmatrix}
										g_{10} & g_{20} \\
										g_{11} & g_{21}
									\end{vmatrix}  \\[0.3cm]
	\pi_{E} &=&	\frac{1}{\Delta}	\begin{vmatrix}
										g_{10} & g_{30} \\
										g_{11} & g_{31}
									\end{vmatrix}, \qquad
	\pi_{F} \;=\; \frac{1}{\Delta}	\begin{vmatrix}
										g_{20} & g_{30} \\
										g_{21} & g_{31}
									\end{vmatrix}  .
\end{eqnarray*}
In these co-ordinates 
\begin{eqnarray*}
	V_{\tau} \w V_{\sigma} &=& \(\pdiff{}{x^{0}} + \dot{\pi}_{I}\,\pdiff{}{\pi_{I}} \) \w  \(\pdiff{}{x^{1}} + \Pi'_{J}\,\pdiff{}{\pi_{J}} \)
\end{eqnarray*}
where $I,J=A,B,C,D,E,F$. We readily calculate
\begin{eqnarray*}
	0 &=& d\Pi(V_{\tau} \w V_{\sigma}) \;=\; d\pi_{A} - \dot{\pi}_{A}\,dx^{0} - \pi'_{A}\,dx^{1} - (\pi'_{B} - \dot{\pi}_{D})dx^{2} - (\pi'_{C} - \dot{\pi}_{E})dx^{3}.
\end{eqnarray*}
But
\begin{eqnarray*}
	d\pi_{A} &=& \dot{\pi}_{A}\,dx^{0} + \pi'_{A}\,dx^{1} + \pdiff{\pi_{A}}{x^{2}}\,dx^{2} + \pdiff{\pi_{A}}{x^{3}}\,dx^{3}
\end{eqnarray*}
so
\begin{eqnarray*}
	0 &=& d\Pi(V_{\tau} \w V_{\sigma}) \;=\;  \pdiff{\pi_{A}}{x^{2}}\,dx^{2} + \pdiff{\pi_{A}}{x^{3}}\,dx^{3} - (\pi'_{B} - \dot{\pi}_{D})dx^{2} - (\pi'_{C} - \dot{\pi}_{E})dx^{3}.
\end{eqnarray*}
Equating to zero each component of the $2-$form gives the two equations
\begin{eqnarray}
	 \label{dPia}	\pdiff{\pi_{A}}{x^{2}}\,dx^{2} - \pi'_{B} + \dot{\pi}_{D} &=& 0 \\[0.2cm]
	 \label{dPib}   \pdiff{\pi_{A}}{x^{3}}\,dx^{3}- \pi'_{C} + \dot{\pi}_{E} &=& 0.
\end{eqnarray}
We now fit the $x^{\mu}$ co-ordinate system into flat spacetime with co-ordinates $y^{\mu}=(y^{+},y^{-},y^{2},y^{3})$ and a ``light cone'' metric
\renewcommand{\arraystretch}{1.1}
\begin{eqnarray*}
	\eta_{\alpha\beta} &=& 	\begin{pmatrix}
								0 & 1 & 0 	& 0 \\
								1 & 0 & 0 	& 0 \\
								0 & 0 & -1 	& 0 \\
								0 & 0 & 0   & -1
							\end{pmatrix}	\begin{array}{c}
												+ \\
												- \\
												2 \\
												3
											\end{array}.			
\end{eqnarray*}\renewcommand{\arraystretch}{1.7}
So 
\begin{eqnarray}\label{StringDsY}
	ds^{2} &=& 2\,dy^{+}\, dy^{-} - (dy^{2})^{2} - (dy^{3})^{2}.
\end{eqnarray}
We try the transformation such that
\begin{eqnarray*}
	y^{+} &=& x^{0} \\
	y^{-} &=& y(x^{0},x^{1}) \\
	y^{2} &=& x^{2} + f(x^{0},x^{1}) \\
	y^{3} &=& x^{3} + g(x^{0},x^{1}) .
\end{eqnarray*}
The string surface will then be given by the equation $x^{2}=x^{3}=0$ or
\begin{eqnarray*}
	y^{2} &=& f(x^{0},x^{1}) \\
	y^{3} &=& g(x^{0},x^{1}) .
\end{eqnarray*}
Now
\begin{eqnarray*}
	dy^{+} &=& dx^{0} \\
	dy^{-} &=& \dot{y}\,dx^{0} + y'\,dx^{1} \\
	dy^{2} &=& dx^{2} + \dot{f}\,dx^{0} + f'\,dx^{1} \\
	dy^{3} &=& dx^{3} + \dot{g}\,dx^{0} + g'\,dx^{1}. 
\end{eqnarray*}
Substituting into (\ref{StringDsY}) gives
\begin{eqnarray}
\begin{split}\label{StringMet}
	ds^{2} 	&\;=\; \(2\dot{y} - (\dot{f}^{2} + \dot{g}^{2})\)\,(dx^{0})^{2} - \( f'^{2} + g'^{2} \)(dx^{1})^{2} - (dx^{2})^{2} - (dx^{3})^{2} \\
			&  \quad + \(2y' - 2\dot{f}\,f' - 2\dot{g}\,g'\)dx^{0}\,dx^{1} - 2\dot{f}\,dx^{2}\,dx^{0} - 2\dot{g}\,dx^{3}\,dx^{0} \\
			&	\quad - 2f'\,dx^{2}\,dx^{1} - 2g'\,dx^{3}\,dx^{1}.
\end{split}
\end{eqnarray}	
We observe that with this transformation we can choose $y$ such that the metric (\ref{StringDsY}) simplifies and can be expressed in terms of $f$ and $g$. The choice
\begin{eqnarray}
	\label{StringDotY}	\dot{y} &=& \frac{1}{2}\(\dot{f}^{2} + \dot{g}^{2} + f'^{2} + g'^{2}\) \\
	\label{StringYDash} 	y' &=& \dot{f}\,f' + \dot{g}\,g'
\end{eqnarray}
yields the metric
\renewcommand{\arraystretch}{1.1}
\begin{eqnarray*}
	g_{\mu\nu} &=& 	\begin{pmatrix}
						f'^{2}+g'^{2} 	& 0 				& -\dot{f} 		& -\dot{g} \\
						0 				& -(f'^{2}+g'^{2}) 	& -f' 			& -g' \\
						-\dot{f}		& -f' 				& -1 			& 0 \\
						-\dot{g}		& -g' 				& 0   			& -1
					\end{pmatrix}	\begin{array}{c}
												x^{0} \\
												x^{1} \\
												x^{2} \\
												x^{3}
											\end{array}.			
\end{eqnarray*}\renewcommand{\arraystretch}{1.7}
Inserting these metric coefficients into (\ref{dPia}) and (\ref{dPib}) we obtain the simple (wave) equations
\begin{eqnarray*}
	\label{StringFWave} \ddot{f} - f'' &=& 0 \\
	\label{StringGWave} \ddot{g} - g'' &=& 0.
\end{eqnarray*}
We see that the conditions (\ref{StringDotY}) and (\ref{StringYDash}) are compatible ($\partial_{\tau}\,\partial_{\sigma}\,y = \partial_{\sigma}\,\partial_{\tau}\,y$) with these equations and that the conditions may be integrated to give $y$ in terms of the independent field degrees of freedom $f$ and $g$ (a new mode will enter here associated with the constant of integration). Given any functions $A,B,C,D$ of a real variable the general solutions to (\ref{StringFWave}) and (\ref{StringGWave})
\begin{eqnarray*}
	f(\sigma,\tau) &=& A(\sigma+\tau) + B(\sigma-\tau) \\
	g(\sigma,\tau) &=& C(\sigma+\tau) + D(\sigma-\tau)
\end{eqnarray*}
can be fitted to the boundary conditions appropriate to the system under consideration. \\

In order to couple an external electromagnetic field to the string manifold we need a $2-$form. The electromagnetic field
\begin{eqnarray*}
	F(x) 	&=& E_{1}\,dx^{1}\w dx^{0} + E_{2}\,dx^{2}\w dx^{0} + E_{3}\,dx^{3} \w dx^{0} \\
			&& \qquad + H_{1}\,dx^{2} \w dx^{3} + H_{2}\,dx^{3} \w dx^{1} + H_{3}\,dx^{1} \w dx^{2} 
\end{eqnarray*}
is such a candidate and by analogy with the electromagnetic coupling of the point particle we investigate the equations that arise from the action
\begin{eqnarray}\label{StringActionF}
	S &=& \int_{V_{\sigma}\w V_{\tau}} (\Pi + q\,F).
\end{eqnarray}
The Euler-Lagrange equations now become
\begin{eqnarray*}
	d(\Pi + q\,F)(V_{\sigma}\w V_{\tau}) &=& 0
\end{eqnarray*}
and the boundary conditions for the open string are
\begin{eqnarray*}
	\int_{V_{\tau}\w \bvec} \Pi + q\,F &=& 0.
\end{eqnarray*}
For a Maxwellian external field $dF=0$, so the Euler-Lagrange equations are unchanged. The boundary conditions may be written 
\begin{eqnarray*}
	\Pi(V_{\tau x}) \pm q\,F(V_{\tau x}) &=& 0
\end{eqnarray*}
or in terms of components
\begin{eqnarray*}
	p_{\sigma}^{\tau}(\sigma_{k},\tau) &=& (-1)^{k}\,q\,F^{\mu}_{\;\nu}(\,\lambda(\sigma_{k},\tau)\,),\,\dot{\lambda}^{\nu}(\sigma_{k},\tau)
\end{eqnarray*}
where $k=0,1$ labels the extremities of the $\sigma$ parameterisation and the $\pm q$ arises since the boundary manifolds are oppositely oriented. The particular interaction (\ref{StringActionF}) has a simple interpretation. By Stokes' theorem
\begin{eqnarray*}
	q\int_{\mathcal{M}_{(2)}} F &=& q\int_{\mathcal{M}_{(2)}} dA \;=\; q\int_{\partial\mathcal{M}_{(2)}} A
\end{eqnarray*}
so the contributions from the spatial boundaries of the string mimic the point particle coupling to the vector potential. Again, the fact that the surface has been assumed orientable automatically produces ``quarks'' at the end points of opposite charge. For the interesting effect that may be conceived when the manifold becomes non-orientable see \cite{NonOrient}. For the closed string of course there are no analogous boundary terms and the system is electrically inert. \\

There is no reason to restrict oneself to Maxwellian field couplings. If $F$ is regarded as a ``tensor potential'' of some new external field (so $dF$ need not vanish) then a non-trivial reparameterisation invariant coupling to the whole string can be achieved. \\

This brief introduction to the classical relativistic string shows how one may employ the exterior calculus to some advantage. The introduction of external field interactions in a consistent way is straightforward and the exploration of gauges an automatic procedure. The conventional approach to canonical quantisation is via the Hamiltonian formalism. Since an extended relativistic system has no unique time parameter there is no unique Hamiltonian. The art is in generating a Hamiltonian which has a tractable structure and that furthermore yields a set of solvable field equations for the manifold. It is for these reasons that the Hamiltonian that generates infinitesimal translations in the global time variable $y^{+}=\tau$ has gained such universal acceptance in the development of a canonical quantisation procedure for the string. \\
\newpage

%%%%%%%%%%%%%%%%%%%%%%%%%%%%%%%%%%%%%%%%%%%%%%%%%%%%%%%%%%%%%%%%%%%%%%%%%%%%%%
\section{The Relativistic Membrane}\label{Sect:RelMembrane}
%%%%%%%%%%%%%%%%%%%%%%%%%%%%%%%%%%%%%%%%%%%%%%%%%%%%%%%%%%%%%%%%%%%%%%%%%%%%%%
The final application will be to the relativistic membrane. As we have noted some local gauge invariant field theories indicate classical solutions that are localised in space and may simulate a confining potential for quark-like fields. Considerable activity has been expended in formulating such ideas into bag model of the hadrons. Since the quantum theory of strings has peculiarities the idea that other localised entities may approximate strings (particularly if they have high angular momenta) in some circumstances but produce a quantitatively distinct theory has attractive possibilities. Such a new approach requires a number of ingredients among which an underlying local gauge invariance is probably essential. However it is of interest to consider in direct analogy with the free relativistic string the classical equation of motion of a $2-$dimensional space-like manifold that has a bounded extent in space. As a relativistic entity it will be described by a $3-$dimensional manifold with a local parameterisation
\begin{eqnarray*}
\begin{split}
	\lambda: \real^{3} &\;\longrightarrow\; \real^{4} \\
		(\tau,\sigma,\rho) &\;\longmapsto\;	x^{\mu} \;=\; \lambda^{\mu}(\tau,\sigma,\rho). 
\end{split}
\end{eqnarray*}
At each point in this manifold we assume that the tangent vectors
\begin{eqnarray*}
	\dot{\lambda}\,\pdiff{}{x^{\mu}}, \quad \lambda'\,\pdiff{}{x^{\mu}}, \quad \bar{\lambda}\,\pdiff{}{x^{\mu}} \qquad \(\text{where}\quad \bar{\lambda}\equiv\pdiff{\lambda^{\mu}}{\rho}\)
\end{eqnarray*}
are oriented with respect to a spacetime frame so that
\begin{eqnarray*}
	\dot{\lambda}^{2} \;\geq\; 0, \qquad \lambda'^{2}\;\leq\; 0 \qquadand \bar{\lambda}^{2}\;\leq\;0. 
\end{eqnarray*}
The action consists of a $3-$form in a $7-$dimensional phase space and working by analogy with the two previous cases we postulate
\begin{eqnarray*}
	\Pi &=& \pi_{\alpha\beta\gamma}\,dx^{\alpha} \w dx^{\beta} \w dx^{\gamma} \\[0.1cm]
	\pi_{\alpha\beta\gamma} &=& \frac{1}{6\Delta}\,\dot{\lambda}_{[\alpha}\,\lambda'_{\beta}\,\bar{\lambda}_{\gamma]}.
\end{eqnarray*}
where $\Delta$ is chosen so that
\begin{eqnarray*}
	\Pi(\wt{\Pi}) &=& 1. 
\end{eqnarray*}
Phase space, therefore, has co-ordinates $x^{\mu},\pi^{\alpha\beta\gamma}$ where the chain is
\begin{eqnarray*}
\begin{split}
	\mathcal{C}_{(3)}: \real^{3} &\;\longrightarrow\; \real^{7} \\
					(\tau,\sigma,\rho) &\;\longmapsto\; \{x^{\mu} \;=\; \lambda^{\mu}(\tau,\sigma,\rho), \pi_{\alpha\beta\gamma}\;=\;\phi_{\alpha\beta\gamma}(\tau,\sigma,\rho) \;\;|\;\; \Pi(\wt{\Pi}) \;=\; 1\}. 
\end{split}
\end{eqnarray*}
As before, introduce phase space tangent vectors 
\begin{eqnarray*}
	V_{\sigma} &\equiv& V_{\sigma x} + V_{\sigma \pi} \;=\; \lambda'^{\alpha}\,\pdiff{}{x^{\alpha}} + \phi'^{\alpha\beta\gamma}\,\pdiff{}{\pi^{\alpha\beta\gamma}} \\[0.2cm]
	V_{\rho} &\equiv& V_{\rho x} + V_{\rho \pi} \;=\; \bar{\lambda}^{\alpha}\,\pdiff{}{x^{\alpha}} + \bar{\phi}^{\alpha\beta\gamma}\,\pdiff{}{\pi^{\alpha\beta\gamma}} \\[0.2cm]
	V_{\tau} &\equiv& V_{\tau x} + V_{\tau \pi} \;=\; \dot{\lambda}^{\alpha}\,\pdiff{}{x^{\alpha}} + \dot{\phi}^{\alpha\beta\gamma}\,\pdiff{}{\pi^{\alpha\beta\gamma}}
\end{eqnarray*}
and write the action for the free relativistic membrane as
\begin{eqnarray}\label{MemAction}
	S &=& \frac{\hbar}{\Omega}\int_{V_{\tau} \w V_{\sigma}\w V_{\rho}} \Pi.
\end{eqnarray}
If the membrane is a closed space-like surface (like a bubble) the variational principle gives the Euler-Lagrange equations
\begin{eqnarray}\label{MemdPi}
	d\Pi(V_{\tau} \w V_{\sigma}\w V_{\rho}) &=& 0
\end{eqnarray}
together with
\begin{eqnarray*}
	P_{\tau} &=& \Pi(V_{\sigma}\w V_{\rho})
\end{eqnarray*}
as the conserved momentum current density $1-$form. In addition to $P_{\tau}$, it is useful to introduce the $1-$forms
\begin{eqnarray*}
	P_{\sigma} &=& \Pi(V_{\rho}\w V_{\tau}) \\[0.1cm]
	P_{\rho} &=& \Pi(V_{\tau}\w V_{\sigma}).
\end{eqnarray*}
In the co-ordinate basis $(x^{\mu},\pi^{\alpha\beta\gamma})$ the covariant equations of motion (\ref{MemdPi}) become after a short calculation
\begin{eqnarray}
\begin{split}\label{MemEOM}
	& \pdiff{}{\tau}\( \frac{\dot{\lambda}^{[\alpha}\,\lambda'^{\beta}\,\bar{\lambda}^{\gamma]}\,\lambda'_{[\gamma}\,\bar{\lambda}_{\beta]} }{\Delta}\) - \pdiff{}{\sigma}\( \frac{\dot{\lambda}^{[\alpha}\,\lambda'^{\beta}\,\bar{\lambda}^{\gamma]}\,\dot{\lambda}_{[\gamma}\,\bar{\lambda}_{\beta]} }{\Delta}\) \\
		& \hspace{6cm} + \pdiff{}{\rho}\( \frac{\dot{\lambda}^{[\alpha}\,\lambda'^{\beta}\,\bar{\lambda}^{\gamma]}\,\dot{\lambda}_{[\gamma}\,\lambda'_{\beta]} }{\Delta}\) \;=\; 0.
\end{split}
\end{eqnarray}
where
\begin{eqnarray*}
	\Delta &=& 	\begin{vmatrix}
					\dot{\lambda}^{2}				&	\dot{\lambda}\cdot\lambda'	&	\dot{\lambda}\cdot\bar{\lambda} \\
					\dot{\lambda}\cdot\lambda'		&	\lambda'^{2}				&	\lambda'\cdot\bar{\lambda} \\
					\dot{\lambda}\cdot\bar{\lambda}	&	\lambda'\cdot\bar{\lambda} 	&	\bar{\lambda}^{2}
				\end{vmatrix}.
\end{eqnarray*}
Because of the general $\tau,\sigma,\rho$ reparameterisation invariance only one of these 4 equations is independent of the others. \\

If the membrane has a boundary then in order to obtain the above Euler-Lagrange equations (\ref{MemdPi}) the displaced manifold must be closed by the addition of a boundary contribution. If the $2-$dimensional manifold swept out by the membrane boundary in spacetime is given by the parameterisation
\begin{eqnarray*}
	x^{\mu} &=& B^{\mu}(\tau,s)
\end{eqnarray*}
then the displaced ``volume'' will be
\begin{eqnarray*}
	\pdiff{}{\tau} \w \pdiff{}{s} \w \pdiff{}{r}
\end{eqnarray*}
where $\pdiff{}{r}$ is an arbitrary vector. The boundary condition becomes
\begin{eqnarray*}
	\Pi\(\pdiff{}{\tau} \w \pdiff{}{s} \) &=& 0.
\end{eqnarray*}
In any particular system of co-ordinates this is a rather complicated (free boundary) condition to impose.\\

The existence of identities follows from vector algebra. The (norm)${}^{2}$ of $P_{\tau}$ is
\begin{eqnarray*}
	P_{\tau}(\wt{P}_{\tau}) &=& \Pi(V_{\sigma x} \w V_{\rho x})\,\wt{\Pi}(\wt{V}_{\sigma x} \w \wt{V}_{\rho x}) \;=\; \frac{R}{\Delta^{2}}
\end{eqnarray*}
where
\begin{eqnarray*}
	\Pi &=& \frac{\wt{V}_{\tau x} \w \wt{V}_{\sigma x} \w \wt{V}_{\rho x}}{\sqrt{(V_{\tau x} \w V_{\sigma x} \w V_{\rho x})(\wt{V}_{\tau x} \w \wt{V}_{\sigma x} \w \wt{V}_{\rho x})  \;}}.
\end{eqnarray*}
But 
\begin{eqnarray*}
	R 	&=& \(\,(\wt{V}_{\tau x} \w \wt{V}_{\sigma x} \w \wt{V}_{\rho x})(V_{\sigma x} \w V_{\rho x})\,\)\[ \,(V_{\tau x} \w V_{\sigma x} \w V_{\rho x})(\wt{V}_{\sigma x} \w \wt{V}_{\rho x}) \,\] \\[0.1cm]
		&=& (\wt{V}_{\tau x} \w \wt{V}_{\sigma x} \w \wt{V}_{\rho x})\left\{\, (V_{\sigma x} \w V_{\rho x}) \w \[ \,(V_{\tau x} \w V_{\sigma x} \w V_{\rho x})(\wt{V}_{\sigma x} \w \wt{V}_{\rho x}) \,\] \,\right\}\\[0.1cm]
		&=& (\wt{V}_{\tau x} \w \wt{V}_{\sigma x} \w \wt{V}_{\rho x})\left\{\, \[\,(V_{\tau x} \w V_{\sigma x} \w V_{\rho x})(\wt{V}_{\sigma x} \w \wt{V}_{\rho x}) \,\] \w (V_{\sigma x} \w V_{\rho x})  \,\right\}.
\end{eqnarray*}
The term in square brackets gives
\begin{eqnarray*}
\begin{split}
	(V_{\tau x} \w V_{\sigma x} \w V_{\rho x})(\wt{V}_{\sigma x} \w \wt{V}_{\rho x}) &\;=\; V_{\tau x}\(\,(V_{\sigma x} \w V_{\rho x})(\wt{V}_{\sigma x} \w \wt{V}_{\rho x})\,\) \\
		&\quad - (V_{\sigma x} \w V_{\rho x})\(\,V_{\tau x}(\wt{V}_{\sigma x} \w \wt{V}_{\rho x})\,\).
\end{split}
\end{eqnarray*}
By exterior multiplication by $V_{\sigma x} \w V_{\rho x}$ this yields
\begin{eqnarray*}
	R &=&  (\wt{V}_{\tau x} \w \wt{V}_{\sigma x} \w \wt{V}_{\rho x})\left\{ \, (V_{\sigma x} \w V_{\rho x})(\wt{V}_{\sigma x} \w \wt{V}_{\rho x})(V_{\tau x} \w V_{\sigma x} \w V_{\rho x}) \, \right\}
\end{eqnarray*}
so 
\begin{eqnarray*}
	P_{\tau}(\wt{P}_{\tau}) - (V_{\sigma x} \w V_{\rho x})(\wt{V}_{\sigma x} \w \wt{V}_{\rho x}) &=& 0.
\end{eqnarray*}
One can also calculate
\begin{eqnarray*}
	P_{\tau}(V_{\sigma x}) &=& \Pi(V_{\sigma x} \w V_{\rho x})(V_{\sigma x}) \;=\; \Pi(V_{\sigma x} \w V_{\rho x} \w V_{\sigma x}) \;=\; 0 \\[0.2cm]
	P_{\tau}(V_{\rho x}) &=& \Pi(V_{\sigma x} \w V_{\rho x})(V_{\rho x}) \;=\; \Pi(V_{\sigma x} \w V_{\rho x} \w V_{\rho x}) \;=\; 0.
\end{eqnarray*}
In a co-ordinate basis
\begin{eqnarray*}
	(V_{\sigma x} \w V_{\rho x})(\wt{V}_{\sigma x} \w \wt{V}_{\rho x}) &=& \lambda'^{2}\,\bar{\lambda}^{2} - (\lambda'\cdot\bar{\lambda})^{2}
\end{eqnarray*}
and the identities may be written
\begin{eqnarray*}
	P_{\tau}^{2} - \lambda'^{2}\,\bar{\lambda}^{2} + (\lambda'\cdot\bar{\lambda})^{2} &=& 0 \\[0.2cm]
	P_{\tau}\cdot\lambda' &=& 0 \\[0.2cm]
	P_{\tau}\cdot\bar{\lambda} &=& 0.
\end{eqnarray*}
In a Hamiltonian formalism in which $P^{\mu}_{\tau}$ is a momentum conjugate to $x^{\mu}$ these are the three primary constraints of the theory. There are other identities that are obtained by repeating the above calculations but with $P_{\tau}$ replaced by $P_{\sigma}$ and $P_{\rho}$ respectively. In this manner we obtain the equations
\begin{eqnarray*}
	P_{\sigma}^{2} - \dot{\lambda}^{2}\,\bar{\lambda}^{2} + (\dot{\lambda}\cdot\bar{\lambda})^{2} &=& 0 \\[0.2cm]
	P_{\sigma}\cdot\dot{\lambda} &=& 0 \\[0.2cm]
	P_{\sigma}\cdot\bar{\lambda} &=& 0
\end{eqnarray*}
and
\begin{eqnarray}
	\label{memPrho} P_{\rho}^{2} - \dot{\lambda}^{2}\,\lambda'^{2} + (\dot{\lambda}\cdot\lambda')^{2} &=& 0 \\[0.2cm]
	\nonumber P_{\rho}\cdot\dot{\lambda} &=& 0 \\[0.2cm]
	\nonumber P_{\rho}\cdot\lambda' &=& 0
\end{eqnarray}
in a co-ordinate basis. Equation (\ref{memPrho}) is of interest if examined together with the boundary condition for the open membrane. If we choose the boundary to be the image of the line $\rho=$constant then $\bar{x}=0$ on the boundary and we can write $V_{\sigma}=V_{\sigma x}$. Then
\begin{eqnarray*}
	\Pi(V_{\tau x} \w V_{\sigma x}) &=& P_{\rho}
\end{eqnarray*}
must be a zero $1-$form on the boundary. However its norm is an element of area on the surface described by the elements $V_{\tau x} \w V_{\sigma x}$ since
\begin{eqnarray*}
	\text{area} &=& \left|V_{\tau x} \w V_{\sigma x}\right|\,d\tau\,d\sigma \;=\; \left|\dot{x}^{2}\,x'^{2} - (\dot{x}\cdot x')^{2}\right|^{\frac{1}{2}}\,d\tau\,d\sigma.
\end{eqnarray*}
Thus we see that the area of the world tube swept out by the open membrane boundary must be null. \\

As in the case of the string, the 4 covariant equations of motion (\ref{MemEOM}) carry redundant information. Choosing a gauge for the relativistic membrane is, however, considerably more difficult than for the string. A $3-$dimensional manifold is not in general conformally flat and unless the membrane has a high degree of symmetry a number of overlapping co-ordinate systems must be considered. Although we can derive an independent equation of motion a gauge choice that yields a simple Hamiltonian is elusive. In all cases that have been studied \cite{GaugeMem} non-linear equations of motion arise with all their attendant problems when one considers quantisation.\\

Let us choose a set of curvilinear co-ordinates $x^{\mu}$ with metric $g_{\mu\nu}$ in spacetime with which to describe the membrane. If the $3-$dimensional manifold under consideration is the co-ordinate ``surface''
\begin{eqnarray*}
	x^{3} &=& 0
\end{eqnarray*}
we shall write
\begin{eqnarray*}
	x^{\mu} &=& \{\,\tau,\sigma,\rho,x^{3}\,\}.
\end{eqnarray*}
The $3-$form $\Pi$ has 4 components so let us write
\begin{eqnarray*}
	\Pi &=& \pi_{A}\,dx^{0} \w dx^{1} \w dx^{2} + \pi_{B}\,dx^{0} \w dx^{1} \w dx^{3} \\
		&& \qquad + \pi_{C}\,dx^{0} \w dx^{2} \w dx^{3} \pi_{D}\,dx^{1} \w dx^{2} \w dx^{3}.
\end{eqnarray*}
In these co-ordinates we can express the components of $\Pi$ in terms of the components of the metric since
\renewcommand{\arraystretch}{1.1}
\begin{eqnarray*}
	\pi_{A} &=& \frac{1}{\Delta}	\begin{vmatrix}		
										\dot{x}_{0}	&	\dot{x}_{1}	&	\dot{x}_{2} \\
										x'_{0}		&	x'_{1}		&	x'_{2} \\
										\bar{x}_{0}	&	\bar{x}_{1}	&	\bar{x}_{2}
									\end{vmatrix}\qquad \text{etc.}
\end{eqnarray*}\renewcommand{\arraystretch}{1.7}
Thus
\renewcommand{\arraystretch}{1.1}
\begin{alignat*}{2}
	\pi_{A} &\;=\; \Delta \qquad  &
	\pi_{B} &=\; \frac{1}{\Delta}	\begin{vmatrix}		
										g_{00}	&	g_{10}	&	g_{30} \\
										g_{01}	&	g_{11}	&	g_{31} \\
										g_{02}	&	g_{12}	&	g_{32}
									\end{vmatrix} \\[0.3cm]
	\pi_{C} &\;=\; \frac{1}{\Delta}	\begin{vmatrix}		
										g_{00}	&	g_{20}	&	g_{30} \\
										g_{01}	&	g_{21}	&	g_{31} \\
										g_{02}	&	g_{22}	&	g_{32}
									\end{vmatrix} \qquad &
	\pi_{D} &\;=\; \frac{1}{\Delta}	\begin{vmatrix}		
										g_{10}	&	g_{20}	&	g_{30} \\
										g_{11}	&	g_{21}	&	g_{31} \\
										g_{12}	&	g_{22}	&	g_{32}
									\end{vmatrix}.
\end{alignat*}
The 3 tangent vectors in phase space simplify to
\begin{eqnarray*}
	V_{\tau} &=& \pdiff{}{x^{0}} + \dot{\Pi}^{K}\,\pdiff{}{\pi^{K}} \\[0.1cm]
	V_{\sigma} &=& \pdiff{}{x^{1}} + \Pi'^{K}\,\pdiff{}{\pi^{K}} \\[0.1cm]
	V_{\rho} &=& \pdiff{}{x^{2}} + \bar{\Pi}^{K}\,\pdiff{}{\pi^{K}} 
\end{eqnarray*}
where $K=A,B,C,D$. \\

The evaluation of (\ref{MemdPi}) proceeds with the aid of the exterior algebra and one obtains the one equation (as the coefficient of $dx^{3}$)
\begin{eqnarray}\label{MemSINGEQ}
	\pdiff{}{x^{3}}\pi_{A} - \pdiff{}{x^{0}}\pi_{D} + \pdiff{}{x^{1}}\pi_{C} - \pdiff{}{x^{2}}\pi_{B} &=& 0.
\end{eqnarray}
A specific gauge is adopted (at least locally) when we choose the metric components $g_{\mu\nu}$, i.e. fit the curvilinear co-ordinates into spacetime. If we contemplate a closed membrane then we might try to describe it any instant of time by the dependence of some radial co-ordinate $R$ on polar angles $\theta\equiv\sigma$ and $\phi\equiv \rho$. Thus if flat spacetime has co-ordinates $y^{0},y^{1},y^{2},y^{3}$ with metric
\begin{eqnarray*}
	ds^{2} &=& (dy^{0})^{2} - (dy^{1})^{2} - (dy^{2})^{2} - (dy^{3})^{2}
\end{eqnarray*}
we relate these to our membrane co-ordinates by the transformation
\begin{eqnarray*}
	y^{0} &=& x^{0} \\
	y^{1} &=& (R+x^{3})\,\sin(x^{1})\,\sin(x^{2}) \\
	y^{2} &=& (R+x^{3})\,\sin(x^{1})\,\cos(x^{2}) \\
	y^{3} &=& (R+x^{3})\,\cos(x^{1}).
\end{eqnarray*}
Then $x^{3}=0$ gives the membrane at any constant $y^{0}=\tau$ in Euclidean $3-$space as
\begin{eqnarray*}
	y^{1}(\tau,\sigma,\rho) &=& R(\tau,\sigma,\rho)\,\sin(\sigma)\,\sin(\rho) \\
	y^{2}(\tau,\sigma,\rho) &=& R(\tau,\sigma,\rho)\,\sin(\sigma)\,\cos(\rho) \\
	y^{3}(\tau,\sigma,\rho) &=& R(\tau,\sigma,\rho)\,\cos(\sigma).
\end{eqnarray*}
Straightforward algebra gives for the metric components in $\{x^{\mu}\}=\{x^{0},x^{3},\theta,\phi\}$ coordinates:
\renewcommand{\arraystretch}{1.1}
\begin{eqnarray*}
	g_{\mu\nu} &=& 	\begin{pmatrix}
						1-\dot{R}^{2}			&	-\dot{R}\,R_{\sigma}				&	-\dot{R}\,R_{\rho}		&	-\dot{R}	\\
						-\dot{R}\,R_{\sigma}	&	-R_{\sigma}^{2} - \underline{R}^{2}	&	-R_{\sigma}\,R_{\rho}	&	-R_{\sigma} \\
						-\dot{R}\,R_{\rho}		&	-R_{\sigma}\,R_{\rho}				&	-R_{\rho}^{2} - \underline{R}^{2}\sin^{2}(\sigma)	&	-R_{\rho} \\
						-\dot{R}				&	-R_{\sigma}							&	-R_{\rho}				& -1
					\end{pmatrix}	\begin{array}{c}
										\tau \\
										\sigma \\
										\rho \\
										x^{3}
									\end{array}
\end{eqnarray*}\renewcommand{\arraystretch}{1.7}
where
\begin{eqnarray*}
	\underline{R} &=& R + x^{3}, \qquad R_{\sigma} \;=\; \pdiff{R}{\sigma}, \qquad R_{\rho} \;=\; \pdiff{R}{\rho}. 
\end{eqnarray*}
The 4 components of $\Pi$ can be explicitly calculated now and (\ref{MemSINGEQ}) yields a partial differential equation for $R(\tau,\sigma,\rho)$. A simple example is a membrane that maintains a spherical symmetry in time: $R_{\sigma}=R_{\rho}=0$. In this case one obtains the equation
\begin{eqnarray*}
	2R\,\(1-\dot{R}^{2}\) + \pdiff{}{\tau}\(\frac{\dot{R}R^{2}}{\sqrt{1-\dot{R}^{2}\;}}\) &=& 0
\end{eqnarray*}
with the first integral
\begin{eqnarray*}
	\frac{R^{2}}{\sqrt{1-\dot{R}^{2}\;}} &=& \text{constant}.
\end{eqnarray*}
This differential equation can be integrated in terms of Jacobi elliptic functions and the radius of the sphere can be expressed as
\begin{eqnarray*}
	R(\tau) &=& \(\frac{4E}{b}\)^{\frac{1}{4}}\,\text{cn}\( \gamma - (4Eb)^{\frac{1}{4}}\tau \left| \frac{1}{\sqrt{2\;}} \right. \)
\end{eqnarray*}
where $E$ and $b$ are constants of the motion and in terms of the complete first order elliptic integral
\begin{eqnarray*}
	\gamma &=& K\(\frac{1}{\sqrt{2\;}}\).
\end{eqnarray*}
This motion is reminiscent of certain pulsing empty universes in general relativity. The spherical system has a tractable Hamiltonian and one can in principle quantise the motion of this pulsing membrane. By electrically charging the membrane uniformly and considering small oscillations about the equilibrium configuration Dirac has examined similar spin-zero excitations \cite{Dirac}.\\

To develop the membrane theory further is beyond the scope of these notes. It would be interesting to calculate the masses of spinning configurations but the non-linear equations appear to demand a numerical attack. The quantum theory has its own set of difficulties although approximate techniques (such as WKB methods) do suggest themselves in a number of situations.\\
\newpage

\addtocontents{toc}{\protect\vspace{10pt}}

%%%%%%%%%%%%%%%%%%%%%%%%%%%%%%%%%%%%%%%%%%%%%%%%%%%%%%%%%%%%%%%%%%%%%%%%%%%%%%
\section*{Conclusions}
\addcontentsline{toc}{section}{\protect\numberline{}Conclusions}
%%%%%%%%%%%%%%%%%%%%%%%%%%%%%%%%%%%%%%%%%%%%%%%%%%%%%%%%%%%%%%%%%%%%%%%%%%%%%%
We have been mainly concerned with the classical relativistic description of extended objects in space-time in terms of exterior forms and integrals over manifolds. This language has a power and  elegance that can be exploited in many ways. We have not discussed symmetries or conservation laws in general but these can also be formulated in a gauge invariant way. The important operation of Hodge duality has not been mentioned although this is of considerable use in discussing physical field theories. The classical de-Rham theories \cite{ExtCalculus} are also destined, in the author's opinion, to play an important role in advancing our understanding of topological currents. We have seen in the examples presented above that much of the complexity of the formalism can be postponed until one needs explicit co-ordinates and even then a judicious choice of co-ordinate system can prove advantageous. There is, of course, no substitute for solving  equations. However, the formalism does suggest generalisation and furthermore is capable of realising such generalisation with a minimum of computational effort. \\

Ultimately the use of conventional exterior forms in physics must be related to the relevance of anti-symmetric tensors in physical theories. It is, however, possible to enlarge the phase space manifolds discussed in these notes to accommodate strings and membranes with intrinsic spin distributions. If spinorial degrees of freedom can be described in this manner it seems possible that internal symmetries may be established in a similar fashion. \\

These notes have concentrated on the ``particle'' aspect of extended objects. A relativistic ``many particle'' picture properly requires a quantum field theory. In particular local gauge invariant field theories have become fashionable of late for good reason. If the exterior calculus can illuminate some of the geometrical and topological properties of such theories and relate them to strings and membranes then its introduction as a tool in particle physics will have been worthwhile.\\

As a supplement to these introductory lectures, the interested reader may wish to consult \cite{FrameBundle_RWT} where a more up-to-date discussion of extremal immersions in curved manifolds can be found. 

\addtocontents{toc}{\protect\vspace{10pt}}

%%%%%%%%%%%%%%%%%%%%%%%%%%%%%%%%%%%%%%%%%%%%%%%%%%%%%%%%%%%%%%%%%%%%%%%%%%%%%%
\section*{Acknowledgements}
\addcontentsline{toc}{section}{\protect\numberline{}Acknowledgements}
%%%%%%%%%%%%%%%%%%%%%%%%%%%%%%%%%%%%%%%%%%%%%%%%%%%%%%%%%%%%%%%%%%%%%%%%%%%%%%
These notes were delivered as a series of NIMROD lectures at the Rutherford Appleton laboratory by the author in February 1976 (RL-76-022). The author would like to acknowledge a helpful discussion with Dr C. T. J. Dodson of the mathematics department at the University of Lancaster during the construction of these notes. Furthermore, the author is grateful to Prof. Bob Bingham for facilitating their retrieval from the RAL archives and to Dr Timothy J. Walton for patiently editing them into a modern \LaTeX\, format.

\newpage
\def\bibsp{-0.2cm}
%%%%%%%%%%%%%%%%%%%%%%%%%%%%%%%%%%%%%%%%
%%%%%%%%%%%%% BIBLIOGRAPHY %%%%%%%%%%%%%
%%%%%%%%%%%%%%%%%%%%%%%%%%%%%%%%%%%%%%%%
\addtocontents{toc}{\protect\vspace{10pt}}
\addcontentsline{toc}{section}{\protect\numberline{}References}

\newpage

\end{document}